\documentclass[lettersize,journal]{IEEEtran}
\usepackage{amsmath,amssymb,amsfonts}
\usepackage{amsthm}
\newtheorem{theorem}{Theorem}
\newtheorem{proposition}{Proposition}
\newtheorem{definition}{Definition}

\usepackage{algorithmic}
\usepackage{algorithm}
\usepackage{array}
\usepackage[caption=false,font=normalsize,labelfont=sf,textfont=sf]{subfig}
\usepackage{booktabs}
\usepackage{textcomp}
\usepackage{stfloats}
\usepackage{graphicx}
\usepackage{url}
\usepackage{verbatim}
\usepackage{cite}
\usepackage{multirow}
\begin{document}

\title{Network Denoising Revisited: A Ricci-Flow-Inspired Graph Diffusion Method}

\author{Ye Fang~and~Chuan-Xian Ren
        % <-this % stops a space
\thanks{The authors are with the School of Mathematics, Sun Yat-sen University,
Guangzhou 510275, China (e-mail: rchuanx@mail.sysu.edu.cn)}}

% The paper headers
\markboth{Journal of \LaTeX\ Class Files,~Vol.~14, No.~8, August~2021}%
{Shell \MakeLowercase{\textit{et al.}}: A Sample Article Using IEEEtran.cls for IEEE Journals}

\maketitle

\begin{abstract}
Networks provide a fundamental representation of relationships among entities. However, real-world networks are often corrupted by noise caused by measurement errors and inherent stochasticity, hindering the discovery of meaningful structure. Most denoising methods rely on similarity-driven diffusion and ignore the non-Euclidean geometry of graphs, where local variations induce heterogeneous information transport. This motivates a geometric revisit of network denoising. In this work, we propose Ricci-Diffusion, a curvature-guided graph diffusion method inspired by Ricci flow. Specifically, Ricci-Diffusion exhibits a Ricci-flow-like evolution, in which relative edge-level curvature modulates local transport in the diffusion kernel and guides edge-weight updates toward a more regular graph geometry. We further provide a theoretical analysis showing that curvature can distinguish graph structures that common similarity-driven diffusion kernels fail to separate, and that curvature induces first-order corrections in one-step diffusion updates. The resulting diffusion process explicitly characterizes transport heterogeneity across local geometries and admits theoretical convergence to a stable denoised network. Results on real-world and synthetic graphs show that curvature-guided updates and curvature homogenization improve structure recovery and downstream performance.
\end{abstract}

\begin{IEEEkeywords}
Network Denoising, Graph Diffusion, Ricci Flow, Ollivier-Ricci Curvature.
\end{IEEEkeywords}

\section{Introduction}
Networks provide a general representation for modeling relationships among entities and have been widely used across diverse fields such as social science, information science, and life science \cite{newman2003structure, koetter2003algebraic, de2010advantages}. In biological networks, weighted protein–protein interaction (PPI) networks encode the physicochemical interaction strength between proteins, and play a central role in identifying functional modules and characterizing cellular organization. In this representation, nodes typically correspond to system components, while weighted edges encode the strength, similarity, or confidence of relationships between entities. Real-world network data are inevitably contaminated by noise arising from measurement errors, environmental perturbations, and intrinsic stochasticity. Such noise may introduce spurious high-confidence interactions, while weak or missing edges can imply genuine relationships. Thus, the true underlying structure of noisy networks may be obscured, their topological properties distorted, and the reliability of structure analysis, functional inference, and downstream tasks significantly degraded. 

Consequently, effective denoising of noisy networks has become an important problem in network analysis. Most network denoising methods aim to suppress spurious connections and enhance meaningful structural relations by re-evaluating or adjusting edge weights. Some methods focus on similarity-based heuristics, leveraging local paths or neighborhood statistics to increase intra-community connectivity, while others adopt probabilistic modeling or causal inference perspectives to assess direct and indirect dependencies between variables \cite{zhou2009predicting, marks2011protein, kaminski2001evaluating}. Among these strategies, graph diffusion methods have emerged as an important class of techniques for network denoising due to their conceptual simplicity, scalability, and strong empirical performance \cite{feizi2013network, yu2023network, wang2018network}. These methods propagate information defined by diffusion kernels along the network structure to update edge weights, thereby suppressing noise and enhancing global structure. However, most existing diffusion-based methods implicitly treat networks as algebraic objects and overlook their inherent geometric properties: at the level of diffusion kernels, they lack an explicit edge-level criterion for assessing the geometric reliability of transport. This limitation is important because real-world networks often exhibit heterogeneous local geometry: densely connected regions may support reliable intra-structure diffusion, whereas sparse, shortcut-like, or bridge-like structures may induce less reliable cross-region transport. For diffusion-based graph denoising, failing to distinguish these geometrically different transport modes can lead to undesired information propagation.

These observations motivate a revisit of the conventional diffusion paradigm for network denoising. Rather than treating denoising as similarity-driven propagation over algebraic kernels, we view it as an information transport process in which edge-level reliability is determined by local graph geometry. Graph curvature provides a natural representation for this purpose, since curvature notions such as Ollivier--Ricci and Forman--Ricci curvature characterize local graph geometry at the edge level \cite{ollivier2007ricci, forman2003bochner}. Therefore, an intuitive approach is to modulate information transport via curvature. Ricci flow further provides a natural mathematical reference for this idea. Ricci flow describes curvature-driven metric evolution, where regions of positive curvature tend to contract while regions of negative curvature expand, progressively uniformizing curvature and revealing the underlying topological structure of a manifold \cite{hamilton1982three, perelman2002entropy}. Discrete analogues of Ricci flow have also been studied on graphs. Related graph-based studies have shown that curvature-guided metric updates can help reveal community structure \cite{ni2019community, jia2026cgdock}.

\begin{figure}[!htbp]
    \centering
    \includegraphics[width=1.0\linewidth]{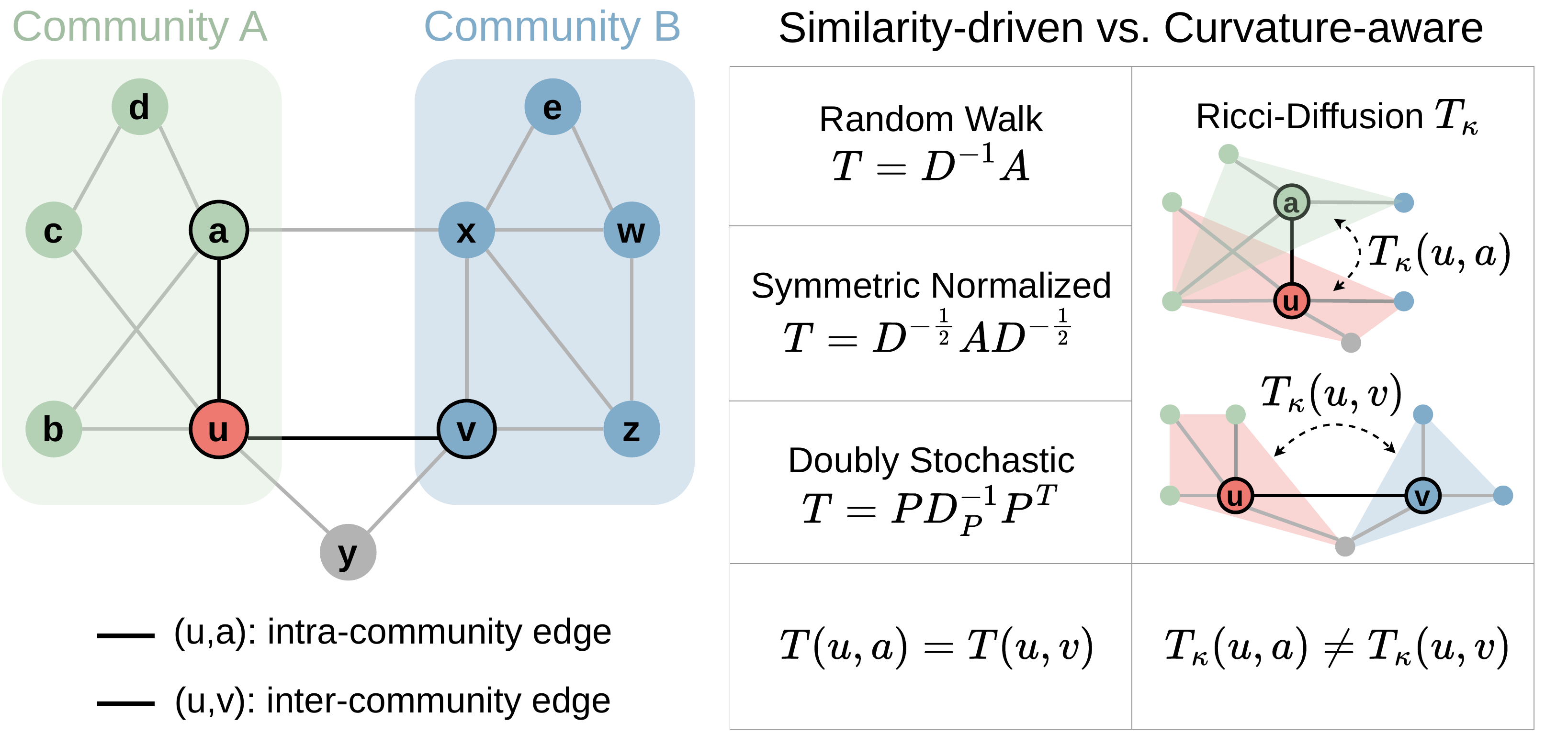}
    \caption{Toy example comparing similarity-driven and curvature-aware diffusion. Common similarity-driven kernels assign identical transport strength to an intra-community edge \((u,a)\) and an inter-community edge \((u,v)\), whereas the curvature-aware Ricci-Diffusion kernel \(T_\kappa\) distinguishes them.}
    \label{fig:toy_example1}
\end{figure}

In this work, we propose \emph{Ricci-Diffusion}, a Ricci-flow-inspired graph diffusion method for network denoising. It explicitly accounts for geometry-induced differences in information transport, yielding denoised networks with clearer geometric structure. By introducing curvature into the diffusion kernel, Ricci-Diffusion can characterize local structures that common similarity-driven diffusion methods fail to separate. Theorem~\ref{thm:kernel_indistinguishability_curvature_separation} establishes the existence of such structures, and Fig.~\ref{fig:toy_example1} provides a representative special case. Specifically, Ricci-Diffusion iteratively updates edge weights using a curvature-aware diffusion kernel in which transport is biased by the relative curvature of an edge within its local neighborhood. This mechanism down-weights unreliable bridge-like transport and promotes structurally coherent propagation, leading to a denoising process with progressively concentrated curvature distributions. Meanwhile, the proposed method admits theoretical convergence guarantees. Experiments on real-world biological networks and controlled synthetic graphs further demonstrate that, by emulating Ricci-flow-like behavior on graphs, Ricci-Diffusion consistently improves performance on downstream tasks.

Our contributions are summarized as follows. 
\begin{itemize}
    \item We revisit the conventional diffusion paradigm for network denoising from a geometric perspective and propose Ricci-Diffusion, a curvature-guided graph diffusion framework inspired by Ricci-flow-like evolution. The method embeds edge-level Ricci curvature into diffusion kernels, enabling local transport to adapt to heterogeneous graph geometry.

    \item We provide a theoretical analysis showing that curvature contributes information beyond similarity-driven propagation. We prove that certain graph structures cannot be distinguished by common diffusion kernels but can be separated by curvature. This curvature further induces first-order geometry corrections to standard diffusion.

    \item We evaluate Ricci-Diffusion on real-world biological networks and controlled synthetic graphs. The results demonstrate improved structure recovery, competitive downstream performance, and empirical Ricci-flow-like behavior through curvature-guided edge-weight evolution and curvature homogenization.
\end{itemize}

\section{Related Work}

Early graph denoising methods aim to recover reliable structures from noisy networks by exploiting structural consistency and statistical dependencies. Local similarity–based approaches infer spurious or missing edges from neighborhood patterns, based on the assumption that meaningful connections preserve local topological coherence \cite{zhou2009predicting}. Complementary work models graphs as noisy observations of latent dependency structures and employs probabilistic or information-theoretic frameworks to disentangle direct interactions from indirect correlations via global consistency constraints \cite{marks2011protein, kaminski2001evaluating, wainwright2008graphical}. While foundational, these methods generally rely on static structural criteria or global inference objectives, rather than modeling denoising as an explicit dynamic process on graphs.

To overcome this limitation, later work formulates graph denoising as a dynamic diffusion process. Network deconvolution removes indirect dependencies by interpreting correlations as multi-hop diffusion and applying spectral inversion \cite{feizi2013network}. Other diffusion-based methods reweight edges via random walks or path-based propagation to enhance community structure \cite{yu2023network}, while Network Enhancement formalizes this idea with a doubly stochastic diffusion operator, yielding a closed-form solution that amplifies spectral eigengaps and improves robustness \cite{wang2018network}. Together, these works establish diffusion as an effective paradigm for graph denoising.

Although diffusion-based methods are effective for graph denoising, they largely rely on similarity-based propagation and lack an explicit geometric perspective on information transport over heterogeneous graph structures. Ricci curvature notions defined on graphs, such as Ollivier–Ricci and Forman–Ricci curvature \cite{ollivier2007ricci, forman2003bochner}, characterize graph geometry through optimal transport or combinatorial constructions and have been exploited in prior work on community detection to reveal network structure \cite{ni2019community}. Inspired by these developments, we incorporate curvature information into the diffusion process to encourage Ricci-flow-like behavior, thereby enabling geometry-aware graph denoising.

\section{Ricci-Diffusion}

\begin{figure*}[htb]
    \centering
    \includegraphics[width=1.0\linewidth]{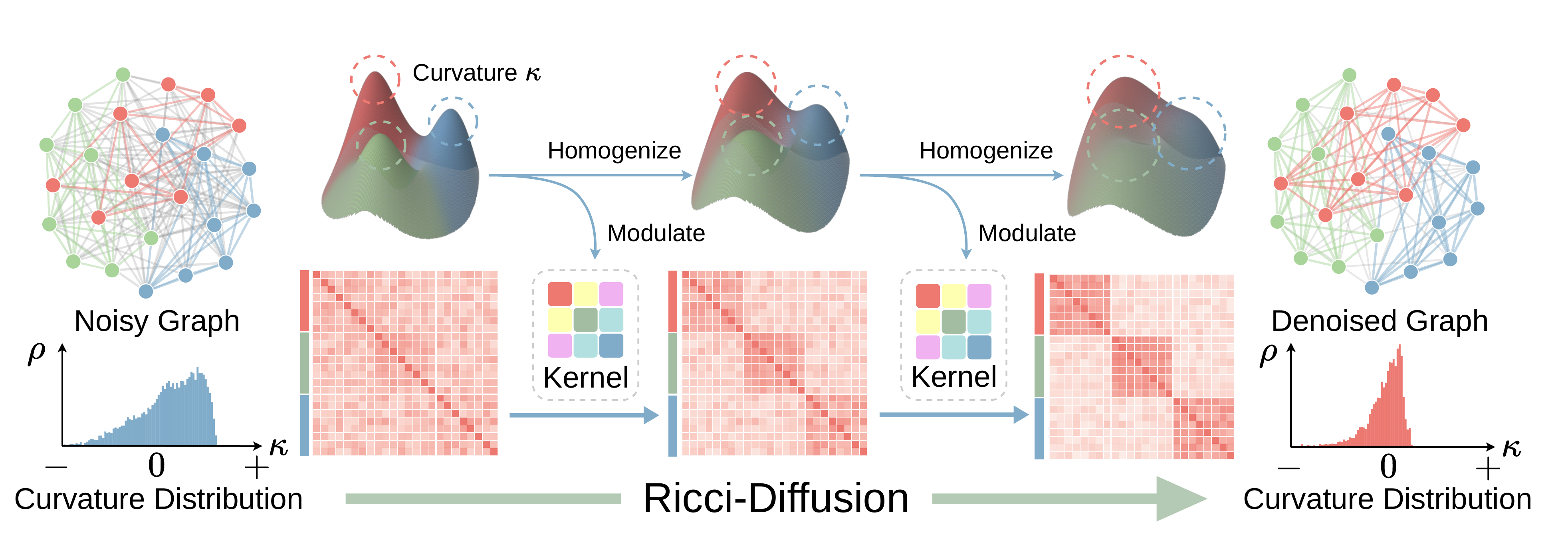}
    \caption{\emph{Illustration of Ricci-Diffusion}. We model graph diffusion--based denoising through a Ricci-flow-inspired graph evolution. Starting from a noisy network, the proposed diffusion method uses curvature-guided transport to strengthen structurally coherent connections while weakening unreliable inter-region connections, yielding a denoised network with a clearer structure with a more concentrated curvature distribution.}
    \label{fig:main_graph}
\end{figure*}

We propose \emph{Ricci-Diffusion}, a graph diffusion method for network denoising. The key idea is to update edge weights through a curvature-aware diffusion kernel. It adapts similarity aggregation and diffusion strength within three-hop neighborhoods according to local geometric variations. This mechanism modulates local transport according to relative edge-level curvature, thereby reducing the influence of unreliable bridge-like edges while promoting structurally coherent propagation, as illustrated in Fig.~\ref{fig:main_graph}. As a result, Ricci-Diffusion exhibits Ricci-flow-like behavior, progressively concentrating curvature distributions and producing denoised networks with a clearer geometric structure.

In the following sections, we first formalize the network setting and introduce a geometric perspective on graph diffusion. We then construct a curvature-aware diffusion kernel that incorporates local geometric information into similarity transport. Building on this kernel, we describe the Ricci-Diffusion process, analyze its generalized diffusion form and convergence behavior, and introduce a dynamic-to-static diffusion strategy. Finally, we provide a theoretical interpretation showing how curvature separates structures missed by similarity-driven kernels and induces first-order corrections in diffusion updates.

\subsection{Network Denoising Problem}

In this problem, we consider an undirected weighted graph $G = (V,E,W)$, where $V$ is the node set with $|V|=n$, $E \subseteq V \times V$ is the edge set, and $W=(w_{ij})\in\mathbb{R}^{n\times n}$ is a symmetric, non-negative weight matrix, i.e., $w_{ij}=w_{ji}\ge 0$. Each weight $w_{ij}$ encodes the similarity or confidence of the relationship between nodes $i$ and $j$. The goal of \emph{network denoising} is to construct a refined graph $\widetilde{G}=(V,E,\widetilde{W})$ whose edge weights better reflect the true relationships and lead to improved performance on downstream tasks.

From a geometric perspective, the graph $G$ can be equipped with a natural metric $d=(d_{ij})_{(i,j)\in E}$ defined by $d_{ij} = f(w_{ij})$, where $f:\mathbb{R}_{\ge 0}\rightarrow\mathbb{R}_{\ge 0}$ is a monotone decreasing function. Under this transformation, the iterative update of edge weights can be equivalently interpreted as an evolution of the induced graph metric.

To formalize the denoising process, we model network denoising as a graph diffusion problem using a generalized diffusion formulation \cite{gasteiger2019diffusion} that captures a broad class of iterative and multi-step behaviors.

\begin{definition}[Generalized Graph Diffusion]
Let $G=(V,E,W)$ denote a weighted graph with $|V|=n$ and let
$T\in\mathbb{R}^{n\times n}$ be a \emph{diffusion kernel}. A generalized diffusion
process is defined as
\[
\widetilde W \;=\; \sum_{k=0}^{\infty}\theta_k\,T^k,
\]
where $\{\theta_k\}_{k\ge0}$ is a sequence of nonnegative coefficients. The
resulting graph is $\widetilde G=(V,E,\widetilde W)$.
\end{definition}

The diffusion kernel determines the behavior of information transportation during denoising. Therefore, we focus on designing a kernel that adapts diffusion dynamics to heterogeneous graph geometry.

\subsection{Curvature-Aware Diffusion Kernel}

\textbf{Geometric Motivation for Graph Diffusion}. Ricci flow \cite{hamilton1982three} offers a geometric perspective that guides the design of a geometry-aware diffusion kernel. Ricci flow evolves a Riemannian metric under curvature, contracting regions of positive curvature and expanding regions of negative curvature, leading to a more regular topological structure.

\begin{definition}[Ricci Flow]
Let $(M, g(t))$ be a smooth manifold with a time-dependent Riemannian metric $g(t)$. 
The \emph{Ricci flow} is a PDE for a Riemannian metric
\[
\frac{\partial}{\partial t} g_{ij}(t) = -2\,\mathrm{Ric}_{ij}(g(t)),
\]
where $\mathrm{Ric}_{ij}$ denotes the Ricci curvature tensor associated with the metric $g(t)$.
\end{definition}

Ricci flow suggests a principle for graph evolution in which curvature participates in metric updates and promotes curvature homogenization. In this sense, Ricci-Diffusion is designed to exhibit Ricci-flow-like behavior: relative edge-level curvature guides local transport during diffusion, and the resulting graph evolution tends to produce a more concentrated curvature distribution and clearer network organization. To realize this principle, we design a curvature-aware diffusion kernel that adapts diffusion dynamics to local graph geometry.

We introduce a curvature-aware diffusion kernel that modulates both similarity aggregation and diffusion intensity based on geometric heterogeneity. The kernel is constructed from a base \textbf{similarity field} that captures local transition structure, together with a \textbf{curvature-guided information transport} mechanism. Through this combination, relative edge-level curvature biases diffusion away from unreliable connections and toward geometrically consistent ones.

\textbf{Local Transition and Similarity Field}. The curvature-aware diffusion kernel is based on a similarity field constructed from multi-hop local transition structure. Given a noisy weighted graph $G=(V,E,W)$, we impose a prior by sparsifying $G$ with a $k$-nearest-neighbors ($k$NN) rule that retains only the top-$k$ strongest connections for each node, yielding a sparse adjacency matrix $W_0$. This operation relies on the assumption that local neighborhood structures remain robust under noise.

Based on the sparsified graph $W_0$, the local transition probability matrix is
computed as
\[
P = D_r^{-1} W_0, \quad D_r = \mathrm{diag}(W_0 \mathbf{1}).
\]
To capture multi-hop transition and higher-order structural consistency, we introduce a similarity field $T_s$ constructed from $P$ via a projection $\Pi_{\mathrm{DSM}}$: 
\[
T_s = \Pi_{\mathrm{DSM}}(P) := P D_P^{-1} P^{\top}, D_P = \mathrm{diag}(P^{\top} \mathbf{1}).
\]
$\Pi_{\mathrm{DSM}}$ maps a row-stochastic matrix $P$ to a \emph{doubly stochastic matrix} (DSM), which satisfies
\[
T_s \mathbf{1} = \mathbf{1}, \quad T_s^{\top} \mathbf{1} = \mathbf{1}, \quad (T_s)_{ij} \ge 0, \quad \forall i,j.
\]
Such DSMs preserve mass in both directions and provide more efficient diffusion \cite{wang2018network, ren2022buresnet, scalzo2023class}.

\textbf{Curvature-Guided Information Transport}. Building on the similarity field, we introduce geometric information to model heterogeneous information transport. \emph{Ollivier–Ricci curvature} (ORC) can measure the contraction and expansion of local neighborhood mass distributions induced by graph geometry. Thus, it serves as a natural signal for geometry-aware diffusion.

To compute ORC, the graph is first endowed with a metric structure by mapping edge weights in $W_0$ to distances $d_{ij} = -\log(w_{ij})$. This construction allows the graph to be viewed as a discrete metric space, providing a geometric basis for characterizing local information transport. Formally, the ORC along an edge $(x,y)\in E$ is defined via the Wasserstein distance as follows.

\begin{definition}[Wasserstein Distance]
Given a metric space $(X,d)$ and two probability measures $\mu$ and $\nu$ on $X$,
the $1$-Wasserstein distance between $\mu$ and $\nu$ is defined as
\[
W(\mu,\nu)
=
\inf_{\gamma\in\Gamma(\mu,\nu)}
\int_{X\times X} d(x,y)\,\mathrm{d}\gamma(x,y),
\]
where $\Gamma(\mu,\nu)$ denotes the set of transportation plans with marginals
$\mu$ and $\nu$.
\end{definition}

\begin{definition}[Ollivier--Ricci Curvature]
Let $m_x$ be a probability measure supported on its neighborhood and $(V,d)$ be a metric graph equipped with probability measures $\{m_x\}_{x\in V}$. The Ollivier--Ricci curvature on an edge $(x,y)$ is given by
\[
\kappa_{xy}
=
1 - \frac{W(m_x,m_y)}{d(x,y)}.
\]
\end{definition}
ORC characterizes the local contraction or expansion of information transport, with positive curvature indicating densely connected regions and negative curvature appearing on sparse or bridge-like edges. This makes ORC a natural geometric signal for modulating transport reliability in diffusion.

To incorporate curvature into the similarity field, we apply an exponential mapping that provides a smooth and nonnegative modulation of transport strength while preserving the underlying transport direction. The resulting curvature-modulated similarity field is then normalized and projected onto the space of doubly stochastic matrices, yielding a stable curvature-aware diffusion kernel.

\begin{definition}[Curvature-Aware Diffusion Kernel] \label{def:kernel}
Given a similarity field $T_s$, an ORC curvature matrix $\kappa$, and a curvature strength parameter \(\eta\ge0\), let
\[
\mathcal{N}_i=\{j:T_s(i,j)>0\}.
\]
The curvature-aware diffusion kernel is defined as
\begin{equation}
\label{eq:curvature_aware_kernel}
\begin{aligned}
T_\kappa(\eta) &= \Pi_{\mathrm{DSM}}\!\big(A_\eta\big),\\
A_\eta(i,j)
&=
\frac{e^{\eta\kappa_{ij}}T_s(i,j)}
{\sum_{\ell\in\mathcal{N}_i}e^{\eta\kappa_{i\ell}}T_s(i,\ell)}.
\end{aligned}
\end{equation}
When \(\eta\) is fixed, we write \(T_\kappa\) for simplicity.
\end{definition}

This kernel provides a principled mechanism for integrating curvature into diffusion, enabling geometry-adaptive information transport and stable mass preservation, and forms the foundation of Ricci-Diffusion as a curvature-guided graph denoising method.

\subsection{Ricci-Diffusion Process}

Based on the curvature-aware diffusion kernel, this section presents the Ricci-Diffusion process, in which edge weights are iteratively updated through the curvature-aware diffusion kernel. We first introduce the iterative update rule and show that the resulting process conforms to the formulation of generalized graph diffusion, while admitting theoretical convergence guarantees. Moreover, a dynamic-to-static diffusion strategy is adopted to improve robustness to noise and enable effective network denoising in practice.

\textbf{One Step Iteration}. Based on the curvature-aware diffusion kernel $T_\kappa$, we define the Ricci-Diffusion process through the following iteration:
\begin{equation} \label{main iter}
    W_{t+1} = \tau\, T_\kappa W_t T_\kappa + (1-\tau)\,T_\kappa,
    \quad \tau\in(0,1).
\end{equation}
The first term corresponds to a random walk within three-hop neighborhoods, where the strength of information transport over the graph is weighted by the curvature-aware diffusion kernel. The second term acts as a regularization term that preserves the backbone of network diffusion and prevents numerical drift during the evolution. As analyzed in Proposition~\ref{prop:ricci_diffusion_geometric_correction}, curvature affects this update through both a direct kernel-level correction and source- and target-side biases in propagated edge weights.

The Ricci-Diffusion process admits a natural interpretation within the generalized graph diffusion framework.

\begin{proposition} \label{prop:generalized_diffusion}
Define the coefficient sequence $\{\theta_k\}$ by
\[
\theta_{2k}=0, \quad \theta_{2k+1}=(1-\tau)\tau^k,\quad k\ge0.
\]
Then the resulting weight matrix $\widetilde{W}$ of Ricci-Diffusion admits a generalized diffusion representation
\[
\widetilde W = \sum_{k=0}^{\infty}\theta_k\,T_{\kappa}^k.
\]
\end{proposition}
The proof of Proposition~\ref{prop:generalized_diffusion} follows by mathematical induction. Since the diffusion kernel in Def.~\ref{def:kernel} is a DSM with spectral radius equal to one, the convergence of Ricci-Diffusion directly follows from the associated matrix geometric series characterized in Proposition~\ref{prop:generalized_diffusion}, leading to the following convergence result.

\begin{proposition}
Suppose $T_{\kappa}$ is a DSM. Then the iteration defined in Eq.~\eqref{main iter} converges to a unique nontrivial fixed point $\widetilde{W}$. Moreover, $\widetilde{W}$ is also a DSM.
\end{proposition}

\begin{algorithm}[t]
\caption{Ricci-Diffusion on Graph}
\label{alg:ricci_diffusion}
\begin{algorithmic}[1]
\STATE \textbf{Input:} Observed noisy graph $G=(V,E,W)$;
$\eta>0$, $\tau\in(0,1)$;
$T_{\mathrm{dyn}}\in\mathbb{N}$;
$q\in\mathbb{N}$;
$k \in \mathbb{N}_+$;
$\varepsilon >0$.
\STATE \textbf{Output:}
Denoised adjacency matrix $\widetilde W$.
\STATE \% Dynamic stage:
\FOR{$t = 0$ {\bfseries to} $T_{\mathrm{dyn}}-1$}
    \IF{$t \bmod q = 0$}
        \STATE Initialize 
        $W_0^{(t)} \leftarrow \mathrm{kNN}(W^{(t)},k)$.
        \STATE Metric 
        $d_{ij}\leftarrow -\log\big((W_0^{(t)})_{ij}\big)$.
        \STATE Local transition matrix
        $P^{(t)} \leftarrow D_r^{-1} W_0^{(t)}$.
        \STATE Similarity transition field
        $T_s^{(t)} \leftarrow \Pi_{\mathrm{DSM}}(P^{(t)})$.
        \STATE Ricci curvature
        $\kappa^{(t)} \leftarrow \mathrm{ORC}(W_0^{(t)};\, d)$.
        \STATE Diffusion kernel
        $T_\kappa^{(t)} \leftarrow \Pi_{\mathrm{DSM}}(A_\eta^{(t)})$ by Eq.~\eqref{eq:curvature_aware_kernel}.
    \ENDIF
    \STATE $W^{(t+1)} \leftarrow \tau\, T_\kappa^{(t)} W^{(t)} T_\kappa^{(t)} + (1-\tau)\, T_\kappa^{(t)}.$
\ENDFOR
\STATE \% Static stage:
\STATE Denote the final dynamic-stage kernel by $T_\kappa^{\mathrm{dyn}}$.
\STATE Update until $\|W^{(t+1)} - W^{(t)}\|_F / \|W^{(t)}\|_F < \varepsilon$:
\[
W^{(t+1)} \leftarrow \tau\, T_\kappa^{\mathrm{dyn}} W^{(t)} T_\kappa^{\mathrm{dyn}} + (1-\tau)\, T_\kappa^{\mathrm{dyn}}.
\]
\STATE \textbf{Output:} $\widetilde W \leftarrow W^{(t)}$.
\end{algorithmic}
\end{algorithm}

\textbf{Dynamic-to-Static Strategy}. In practice, the diffusion kernel can be built directly from the observed noisy network. Instead, we adopt a \emph{dynamic-to-static} strategy to obtain a more reliable curvature-aware kernel.

In the \emph{dynamic stage}, Ricci-Diffusion iteratively updates edge weights and periodically recomputes curvature and the diffusion kernel. This allows the transport geometry to adapt to the evolving graph while avoiding excessive recomputation.

After the dynamic stage, we fix the resulting kernel and enter the \emph{static stage}, where the diffusion process is continued until convergence. This reduces computational cost and preserves the theoretical tractability. The complete algorithm is given in Alg.~\ref{alg:ricci_diffusion}.

\subsection{Theoretical Interpretation of Ricci-Diffusion}

This subsection provides a mechanistic interpretation of Ricci-Diffusion. Theorem~\ref{thm:kernel_indistinguishability_curvature_separation} formalizes the special case illustrated in Fig.~\ref{fig:toy_example1}: for a constructed graph family, common similarity-driven diffusion kernels assign identical transport strength to two edges with different geometric roles, whereas curvature separates them. This separation supplies the local geometric signal needed by Ricci-Diffusion. Proposition~\ref{prop:ricci_diffusion_geometric_correction} then characterizes how this signal affects a one-step update of Ricci-Diffusion through a first-order correction. The detailed proofs are provided in Appendix~\uppercase\expandafter{\romannumeral 1}.

\begin{theorem}[Similarity-kernel limitation and curvature separation]
\label{thm:kernel_indistinguishability_curvature_separation}
There exists a graph family \(\mathcal{M}=\{G_k\}_{k\ge 1}\) with the following property. For any \(G_k\in\mathcal{M}\), let \(A\) be its adjacency matrix, let \(D=\mathrm{diag}(A\mathbf 1)\), and set \(P=D^{-1}A\). Let \(D_P=\mathrm{diag}(P^\top\mathbf 1)\). Denote
\[
\mathcal{T}_{\rm sim}
=
\left\{
P,\;
D^{-1/2}AD^{-1/2},\;
PD_P^{-1}P^\top
\right\}.
\]
Then \( \exists u,a,v\in V(G_k)\) such that
\[
(u,a),(u,v)\in E(G_k),
\]
satisfying
\[
(T_{\rm sim})_{ua}=(T_{\rm sim})_{uv},
\quad
\forall T_{\rm sim}\in\mathcal{T}_{\rm sim},
\]
but
\[
\kappa_{ua}-\kappa_{uv}
=
\frac{2k}{k+3}
>0.
\]
\end{theorem}

This curvature separation provides the edge-level signal missing from the similarity-driven kernels. When this signal is incorporated into the diffusion kernel, its effect on one-step edge-weight evolution can be characterized by the following first-order expansion.

\begin{proposition}[Curvature-induced first-order correction]
\label{prop:ricci_diffusion_geometric_correction}
Let \(T_s\in\mathbb{R}_{\ge 0}^{n\times n}\).
For the parameterized kernel in Def.~\ref{def:kernel}, let \(T_\eta=T_\kappa(\eta)\) and \(T_0=T_\kappa(0)\). Define
\[
\bar{\kappa}_i
=
\sum_{\ell\in\mathcal{N}_i}T_s(i,\ell)\kappa_{i\ell},~
(B_\kappa)_{ij}
=
\begin{cases}
T_s(i,j)(\kappa_{ij}-\bar{\kappa}_i), & j\in\mathcal{N}_i,\\
0, & j\notin\mathcal{N}_i.
\end{cases}
\]
Let
\[
\dot T_0=\mathrm{D}\Pi_{\mathrm{DSM}}[T_s](B_\kappa),
\]
where \(\mathrm{D}\Pi_{\mathrm{DSM}}[T_s]\) denotes the Fréchet derivative of \(\Pi_{\mathrm{DSM}}\) at \(T_s\).
For \(W\in\mathbb{R}_{\ge 0}^{n\times n}\) and \(\tau\in(0,1)\), define
\[
W_\eta^+
=
\tau T_\eta W T_\eta+(1-\tau)T_\eta,
\quad
W_{\eta=0}^+
=
\tau T_0WT_0+(1-\tau)T_0.
\]
Then, as \(\eta\to 0\),
\[
W_\eta^+
=
W_{\eta=0}^+
+
\eta\mathcal{G}_\kappa(W,T_0)
+
O(\eta^2),
\]
where
\[
\mathcal{G}_\kappa(W,T_0)
=
(1-\tau)\dot T_0
+
\tau
\left(
\dot T_0WT_0+T_0W\dot T_0
\right).
\]
\end{proposition}

Proposition~\ref{prop:ricci_diffusion_geometric_correction} explains how the curvature signal identified above enters a one-step diffusion update. Since \((B_\kappa)_{ij}=T_s(i,j)(\kappa_{ij}-\bar{\kappa}_i)\), the correction depends on the relative curvature of an edge with respect to its local neighborhood, rather than on its absolute curvature value alone. Under a small-\(\eta\) expansion, the curvature-modulated update \(W_\eta^+\) equals the baseline update \(W_{\eta=0}^+\) plus the first-order correction \(\eta\mathcal{G}_\kappa(W,T_0)\). In this correction, \((1-\tau)\dot T_0\) gives the direct kernel-level adjustment, while \(\tau\dot T_0WT_0\) and \(\tau T_0W\dot T_0\) describe source-side and target-side curvature biases in the propagated edge-weight update. This shows that Ricci-Diffusion incorporates edge-level geometric information into the diffusion dynamics: curvature guides local weight updates, which is the mechanism underlying its Ricci-flow-like tendency toward curvature homogenization.

\begin{table*}[htb]
\caption{Gene function prediction performance on tissue-specific gene interaction networks. We report AUROC for 16 evaluated tissues in the format mean (std). Higher AUROC indicates stronger gene–function associations supported by the denoised network.}
\label{tissue}
\centering
\renewcommand{\arraystretch}{0.95}
\begin{tabular}{c c c c c c c}
\toprule
Tissue & Raw & ND & NE & NR & RD-Sta & RD-Dyn\\
\midrule
blood plasma& 0.707 (0.041) & \underline{0.735 (0.063)} & 0.731 (0.051) & 0.675 (0.034) & \textbf{0.739 (0.052)} & 0.733 (0.061)\\
blood platelet& 0.570 (0.023) & 0.617 (0.013) & 0.735 (0.050) & 0.551 (0.017) & \underline{0.759 (0.038)} & \textbf{0.763 (0.035)}\\
blood& 0.709 (0.078) & 0.693 (0.072) & \textbf{0.718 (0.065)} & 0.697 (0.080) & \underline{0.702 (0.079)} & 0.689 (0.081)\\
blood vessel& 0.669 (0.049) & 0.697 (0.052) & 0.733 (0.052) & 0.649 (0.052) & \underline{0.758 (0.044)} & \textbf{0.762 (0.041)}\\
b lymphocyte& 0.606 (0.126) & 0.666 (0.096) & 0.751 (0.105) & 0.591 (0.122) & \underline{0.776 (0.090)} & \textbf{0.784 (0.084)}\\
bone& 0.510 (0.043) & 0.491 (0.035) & \textbf{0.558 (0.047)} & 0.490 (0.041) & 0.527 (0.048) & \underline{0.539 (0.047)}\\
brain& 0.569 (-) & \textbf{0.662 (-)} & 0.648 (-) & 0.571 (-) & 0.651 (-) & \underline{0.659 (-)}\\
central nervous system& 0.619 (0.057) & 0.688 (0.061) & 0.686 (0.027) & 0.617 (0.048) & \textbf{0.695 (0.012)} & \underline{0.690 (0.018)}\\
epidermis& 0.549 (0.024) & \textbf{0.622 (0.026)} & \underline{0.606 (0.065)} & 0.542 (0.026) & 0.571 (0.060) & 0.554 (0.051)\\
heart& 0.528 (0.108) & 0.603 (0.078) & 0.682 (0.100) & 0.504 (0.107) & \underline{0.681 (0.100)} & \textbf{0.692 (0.098)}\\
lymphocyte& 0.700 (0.073) & 0.741 (0.058) & 0.782 (0.061) & 0.681 (0.072) & \underline{0.793 (0.058)} & \textbf{0.801 (0.057)}\\
natural killer cell& 0.628 (0.113) & 0.624 (0.093) & 0.631 (0.097) & 0.622 (0.109) & \underline{0.630 (0.097)} & \textbf{0.636 (0.093)}\\
nervous system& 0.673 (0.069) & 0.668 (0.065) & 0.706 (0.050) & 0.654 (0.066) & \underline{0.716 (0.050)} & \textbf{0.718 (0.046)}\\
neuron& 0.564 (0.061) & 0.552 (0.058) & 0.608 (0.053) & 0.512 (0.051) & \underline{0.597 (0.056)} & \textbf{0.610 (0.052)}\\
skeletal muscle& 0.432 (0.070) & 0.546 (0.049) & 0.559 (0.042) & 0.409 (0.069) & \textbf{0.583 (0.041)} & \underline{0.579 (0.048)}\\
t lymphocyte& 0.698 (0.061) & 0.737 (0.050) & 0.769 (0.066) & 0.679 (0.060) & \underline{0.787 (0.059)} & \textbf{0.799 (0.058)}\\
\bottomrule
\end{tabular}
\end{table*}

\section{Experiments}
We systematically evaluate Ricci-Diffusion as a general-purpose network denoising module from four complementary perspectives. First, we assess its downstream utility on three biologically motivated network scenarios, including tissue-specific gene function prediction, chromatin interaction (Hi-C) network denoising for TAD detection, and fine-grained species identification based on species similarity networks. These tasks share a common dependence on network fidelity, where edges encode meaningful biological relationships and effective denoising is expected to benefit subsequent analyses. Second, we examine the denoising dynamics through edge-wise curvature evolution to assess whether Ricci-Diffusion exhibits Ricci-flow-like behavior. Third, we conduct ablation and sensitivity studies to isolate the contribution of curvature modulation from the diffusion backbone. Finally, we use controlled synthetic graphs with known structural ground truth to directly evaluate structure recovery and downstream classification.

For the real-world denoising tasks, we compare Ricci-Diffusion with representative diffusion-based methods, including Network Deconvolution (ND) \cite{feizi2013network}, Network Enhancement (NE) \cite{wang2018network}, and Network Refinement (NR) \cite{yu2023network}. For the synthetic experiments, we further include BORF \cite{nguyen2023revisiting} and GSR \cite{zhao2023self} as additional baselines.

Ricci-Diffusion supports two iteration strategies, namely a dynamic variant (\textbf{RD-Dyn}), which adopts a dynamic-to-static strategy, and a static variant (\textbf{RD-Sta}), which operates without the dynamic stage. Both RD-Sta and RD-Dyn are reported when computationally feasible. The \textbf{best}  results are highlighted in bold, and the \underline{second-best} results are underlined.

Detailed dataset construction, parameter settings, and evaluation protocols are provided in Appendix~\uppercase\expandafter{\romannumeral 2}.

\subsection{Gene Function Prediction on Tissue Networks}

We evaluate network denoising on tissue-specific gene interaction networks \cite{greene2015understanding} through gene function prediction. Denoised networks are assessed by AUROC under the standard gene-function prediction protocol.

\begin{figure*}[t]
    \centering
    \includegraphics[width=1\linewidth]{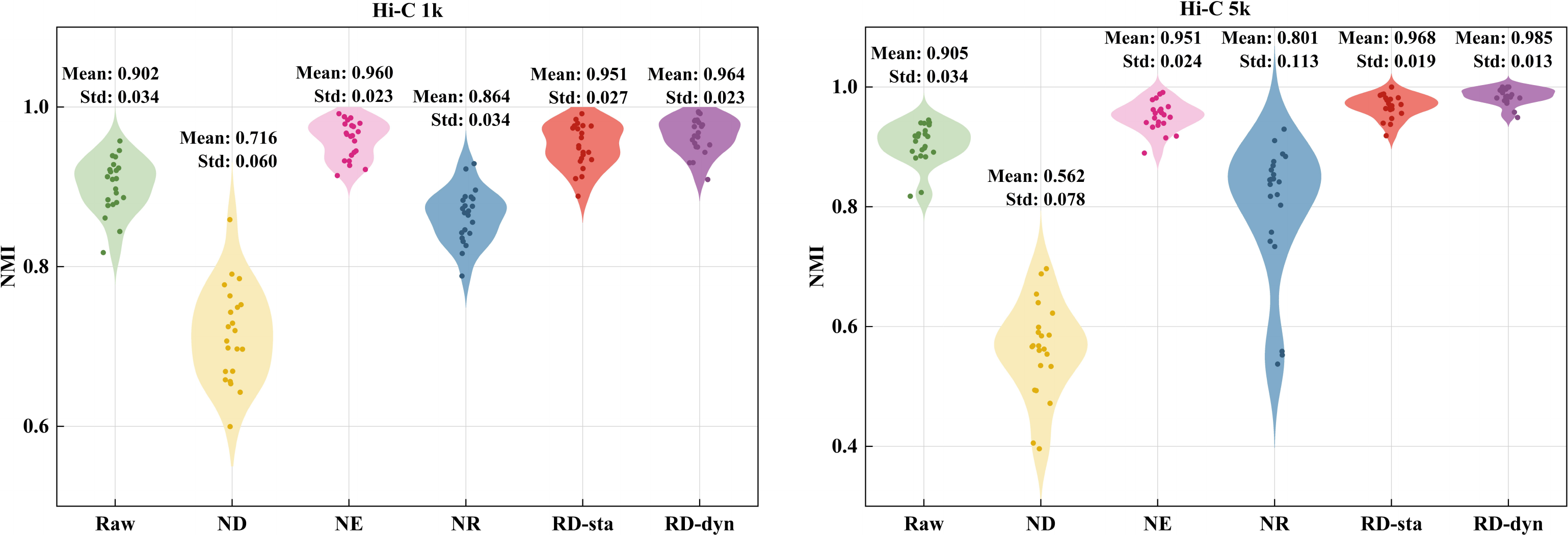}
    \caption{TAD detection performance on Hi-C networks at 1\,k (left) and 5\,k (right) resolutions. Violin plots show the distribution of normalized mutual information (NMI) scores obtained by applying Louvain community detection to raw and denoised Hi-C networks. Each point corresponds to one chromosome. Higher NMI indicates better agreement with the reference structure.}
    \label{fig:Hi-C_results}
\end{figure*}

Table~\ref{tissue} reports AUROC scores for gene function prediction across 16 tissue-specific networks. RD-Sta consistently improves AUROC over the raw networks across all tissues, with particularly pronounced gains observed in blood platelet (0.759 vs.\ 0.570) and b lymphocyte (0.776 vs.\ 0.606), indicating enhanced recovery of biologically meaningful signals in diverse tissue contexts. RD-Dyn further improves performance on several tissues, indicating that dynamic kernel updates can provide additional benefits when the computational cost is acceptable. Compared with NR, a recent graph diffusion method for network denoising, RD-Sta achieves consistent performance gains across most tissues, with particularly large improvements observed in blood platelet (0.759 vs.\ 0.551), b lymphocyte (0.776 vs.\ 0.591), and skeletal muscle (0.583 vs.\ 0.409). Compared with NE, which employs a diffusion kernel with a similar DSM structure but without curvature guidance, RD-Sta attains higher AUROC on 11 out of 16 tissues and remains competitive on the remaining cases, providing evidence that incorporating curvature into the diffusion process has a positive impact on functional information transport.

\subsection{Hi-C Network Denoising for TAD Detection}

We evaluate whether network denoising improves topologically associating domain (TAD) detection from Hi-C contact networks \cite{rao20143d, schmitt2016genome}. We use Hi-C data from the GM12878 cell line at 1\,k and 5\,k resolutions, considering all autosomes. Denoised networks are assessed by normalized mutual information (NMI) between detected communities and reference domain structures.

Fig.~\ref{fig:Hi-C_results} reports NMI scores for TAD detection at 1\,k and 5\,k resolutions. RD-Sta and RD-Dyn consistently outperform NR and ND at both resolutions. Compared with NE, RD-Dyn is slightly higher at 1\,k resolution, while both RD variants show larger gains at 5\,k resolution. This resolution-dependent improvement indicates that when Hi-C networks become sparser and more affected by noise, Ricci-Diffusion is more effective at preserving and recovering faithful network structure than similarity-driven diffusion. Moreover, RD-Dyn consistently outperforms RD-Sta across both resolutions, indicating that dynamically adapting the diffusion kernel to the evolving graph geometry better captures the intrinsic network backbone and improves TAD detection accuracy.

\begin{figure*}[htb]
    \centering
    \includegraphics[width=1\linewidth]{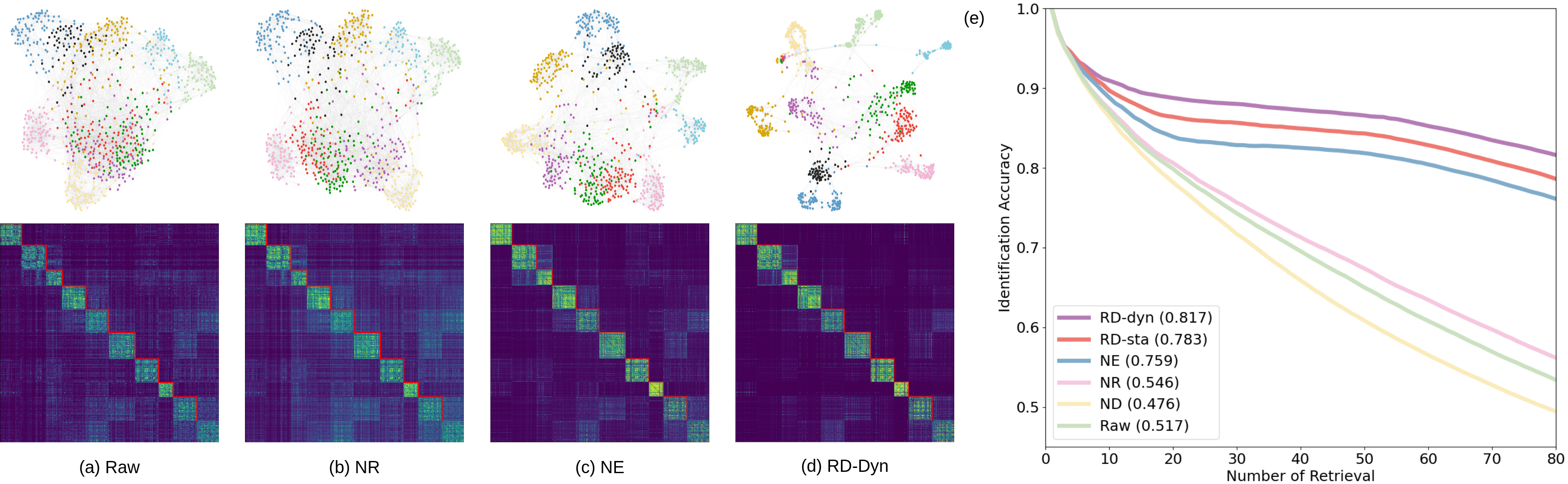}
    \caption{Fine-grained species identification on the Leeds Butterfly dataset. (a–d) Visualization of the image networks after denoising using Raw, NR, NE, and RD-Dyn, respectively, where node colors indicate species labels. RD-Dyn produces clearer and more compact community structures with reduced inter-species connections. (e) Identification accuracy as a function of the number of retrieved neighbors for different network denoising methods. Values in parentheses denote the mean identification accuracy averaged over all query images.}
    \label{fig:butterfly_all}
\end{figure*}

\begin{figure*}[htb]
  \centering

  \begin{minipage}[t]{0.32\textwidth}
    \centering
    \includegraphics[width=\linewidth]{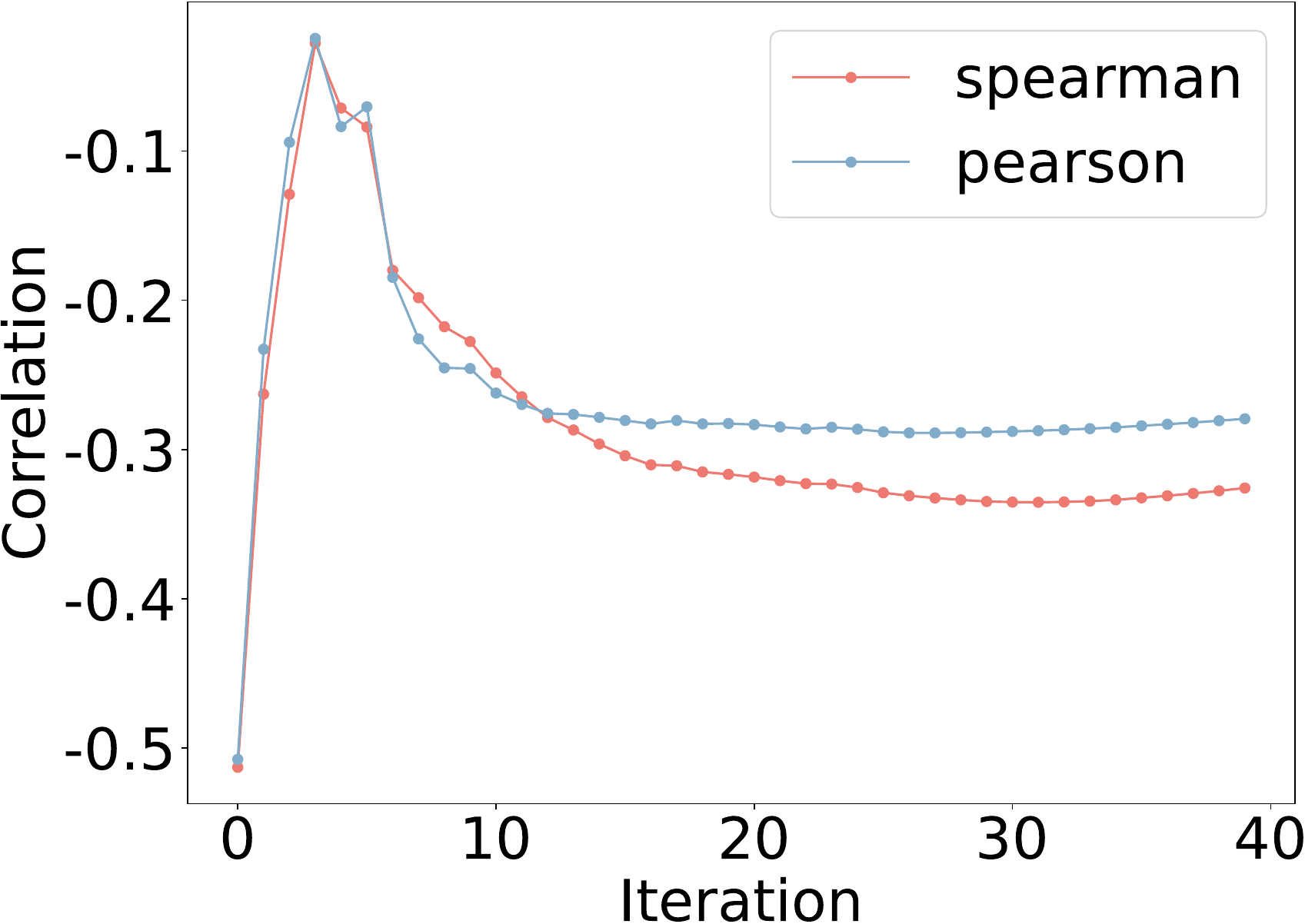}\\[-1mm]
    \small (a)
  \end{minipage}\hfill
  \begin{minipage}[t]{0.33\textwidth}
    \centering
    \includegraphics[width=\linewidth]{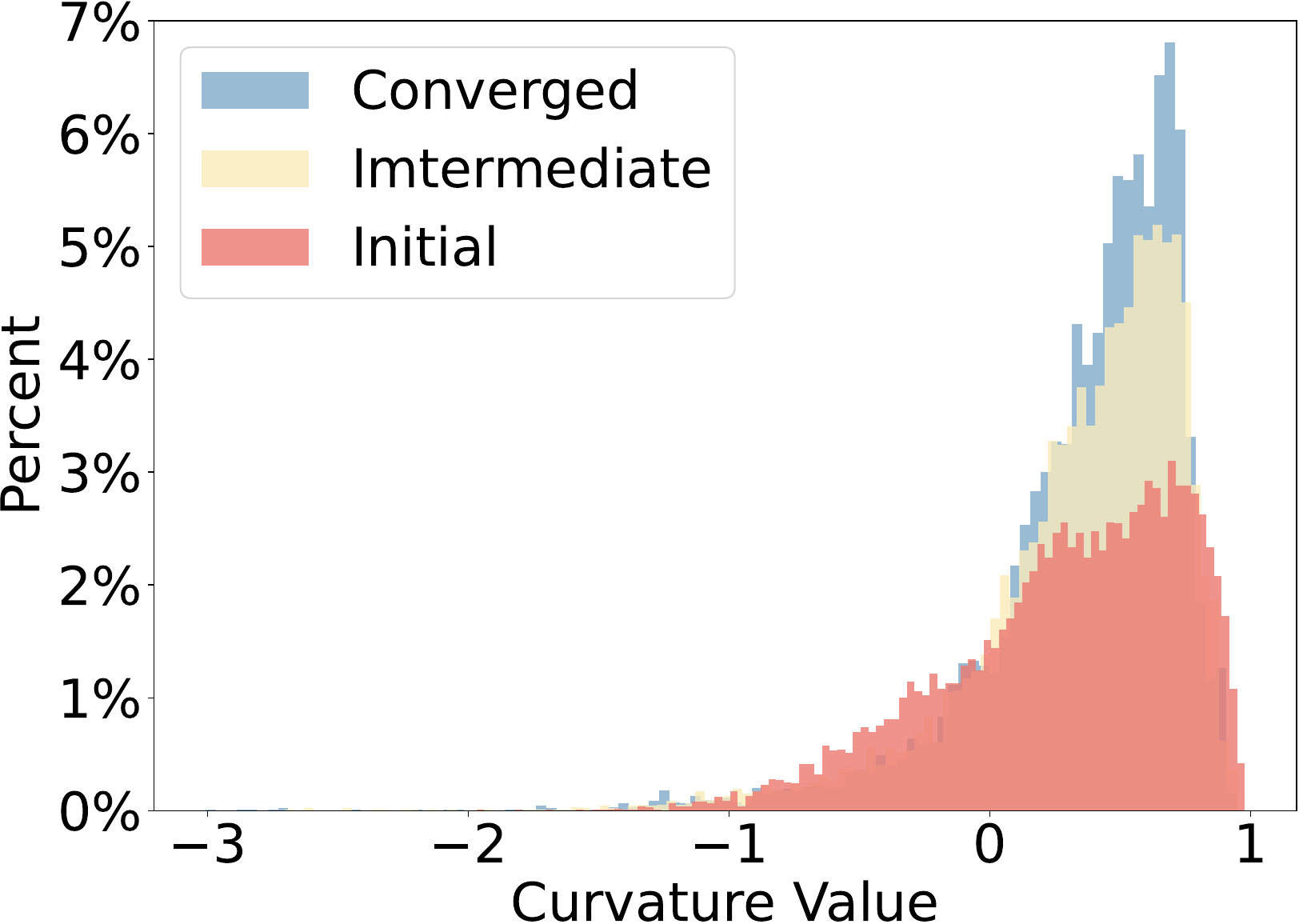}\\[-1mm]
    \small (b)
  \end{minipage}\hfill
  \begin{minipage}[t]{0.32\textwidth}
    \centering
    \includegraphics[width=\linewidth]{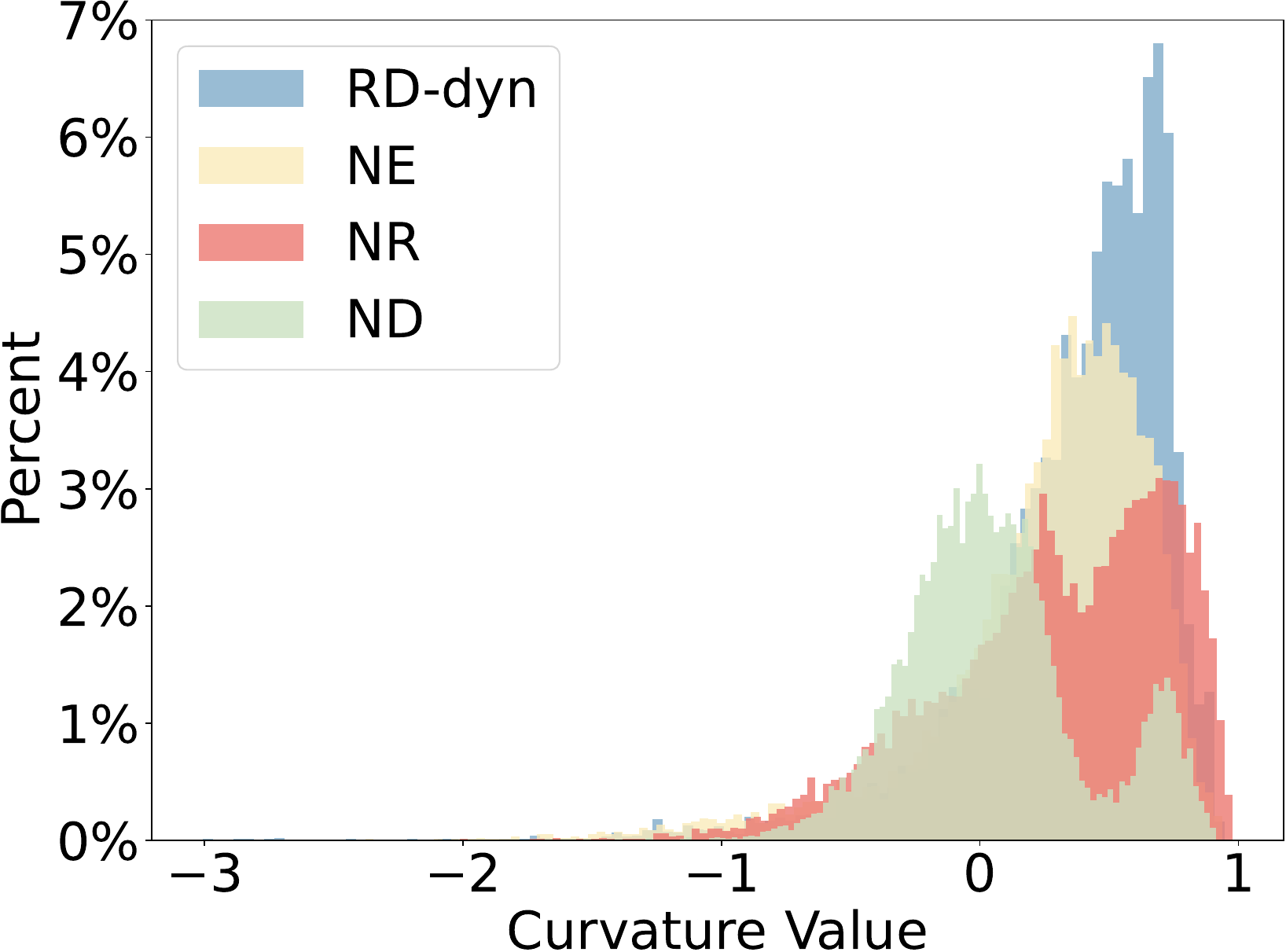}\\[-1mm]
    \small (c)
  \end{minipage}

  \caption{Geometric evidence of Ricci-flow-like behavior in Ricci-Diffusion. 
  (a) Spearman and Pearson correlations between relative edge-weight updates 
  \(r^{(t)}_{ij}\) and ORC \(\kappa^{(t)}_{ij}\) along the diffusion trajectory, showing a persistent negative correlation. 
  (b) Evolution of curvature distributions from initial to converged stages, demonstrating progressive concentration during diffusion. 
  (c) Final curvature distributions produced by different methods, where more concentrated distributions correspond to a clearer geometric structure.}
  \label{fig:ricciflow_like}
\end{figure*}

\subsection{Fine-Grained Species Identification}

We evaluate network denoising methods on fine-grained species identification using the Leeds Butterfly dataset \cite{wang2009learning}, which contains 832 images from 10 species. Denoising performance is assessed through an image retrieval task, where accuracy is measured by the proportion of same-species images among top-ranked neighbors.

Fig.~\ref{fig:butterfly_all} summarizes the identification accuracy achieved by different methods. RD-Dyn achieves the highest mean retrieval accuracy of 0.817, followed by RD-Sta at 0.783, both significantly exceeding NE (0.759), NR (0.546), Raw (0.517), and ND (0.476). These gains are stable across different numbers of retrieved neighbors, demonstrating that curvature-guided diffusion more effectively preserves intra-species similarity while suppressing spurious inter-species connections. Beyond retrieval accuracy, visualization reveals clear structural differences among methods. Compared with Raw (Fig.~\ref{fig:butterfly_all}a), NR (Fig.~\ref{fig:butterfly_all}b), and NE (Fig.~\ref{fig:butterfly_all}c), RD-Dyn (Fig.~\ref{fig:butterfly_all}d) yields markedly cleaner community separation, characterized by compact species-specific clusters and strongly suppressed inter-community connections. This structural sharpening is consistent with the intended Ricci-flow-like behavior: \textit{edges with more negative curvature tend to be suppressed, while positively curved intra-community edges are preserved or reinforced, driving the network toward a clearer organization}.

\subsection{Geometric Validation of Ricci-Flow-Like Behavior}

To provide geometric evidence of Ricci-flow-like behavior in Ricci-Diffusion, we analyze the coupling between edge-weight evolution and Ollivier--Ricci curvature on the Leeds Butterfly dataset. Specifically, we measure the relative edge-weight change $r^{(t)}_{ij}= \log w^{(t)}_{ij} - \log w^{(t+1)}_{ij}$. We then correlate \(r^{(t)}_{ij}\) with the corresponding curvature \(\kappa^{(t)}_{ij}\).
Fig.~\ref{fig:ricciflow_like}a reports the Spearman and Pearson correlations between the relative edge-weight updates $r^{(t)}_{ij}$ and the corresponding curvatures $\kappa^{(t)}_{ij}$ across iterations. Both correlations remain consistently negative, indicating that edges with more negative curvature tend to undergo stronger weight suppression, while positively curved edges are more likely to be preserved. Fig.~\ref{fig:ricciflow_like}b shows that as diffusion proceeds, the curvature distribution progressively concentrates. Fig.~\ref{fig:ricciflow_like}c compares the final curvature distributions produced by different denoising methods, where RD-Dyn yields the most concentrated distribution. These results provide empirical evidence that Ricci-Diffusion exhibits Ricci-flow-like behavior on graphs, in which curvature is correlated with edge-weight updates and the curvature distribution becomes more concentrated during denoising.

\subsection{Ablation Study on Curvature Modulation}

We conduct an ablation study on the Butterfly dataset to isolate the effect of curvature modulation from the diffusion backbone. We compare ORC, FRC, random signals, degree-based proxies, and the no-curvature setting \((\eta=0)\), where curvature modulation is removed while the same diffusion backbone is retained. Extended sensitivity analyzes are provided in the Appendix~\uppercase\expandafter{\romannumeral 4}.

\begin{table}[t]
\centering
\renewcommand{\arraystretch}{0.9}
\caption{Ablation study on the Butterfly dataset.}
\label{tab:ablation_curvature}
\begin{tabular}{lcc}
\toprule
\textbf{Curvature signal} & \textbf{RD-Dyn} & \textbf{RD-Sta} \\
\midrule
No curvature (\(\eta=0\)) & 0.789 & 0.759 \\
\midrule
Random signal & 0.573 & 0.556 \\
Degree & 0.709 & 0.691 \\
\midrule
FRC, \(\eta=1\) & 0.808 & 0.773 \\
FRC, \(\eta=2\) & 0.806 & 0.775 \\
FRC, \(\eta=3\) & 0.801 & 0.776 \\
\midrule
ORC, \(\eta=1\) & 0.817 & 0.783 \\
ORC, \(\eta=2\) & 0.824 & 0.789 \\
ORC, \(\eta=3\) & \textbf{0.825} & \textbf{0.794} \\
\bottomrule
\end{tabular}
\end{table}

Table~\ref{tab:ablation_curvature} shows that curvature modulation improves over the pure diffusion backbone. Removing curvature by setting \(\eta=0\) reduces the performance of both RD-Dyn and RD-Sta, whereas ORC-based modulation improves RD-Dyn from \(0.789\) to \(0.825\) and RD-Sta from \(0.759\) to \(0.794\). Replacing curvature with random or degree-based signals leads to weaker performance, indicating that the gain is not explained by arbitrary perturbations or simple structural proxies. FRC also improves over the no-curvature setting in most cases, while ORC gives the strongest results, suggesting that transport-based curvature better captures the local geometric heterogeneity relevant to diffusion-based denoising.

\begin{table*}[t]
\centering
\caption{Controlled synthetic denoising results. The metric is inv-SNR. Lower is better.}
\label{tab:synthetic_inv_snr}
\scriptsize
\renewcommand{\arraystretch}{0.95}
\setlength{\tabcolsep}{4.0pt}
\resizebox{\textwidth}{!}{
\begin{tabular}{cccccccccc}
\toprule
\multirow{2}{*}{\textbf{Dataset}}
& \multirow{2}{*}{\textbf{Noise}}
& \multirow{2}{*}{\textbf{Raw}} 
& \multicolumn{3}{c}{\textbf{Diffusion baselines}} 
& \textbf{Curv.} 
& \textbf{Learn.} 
& \multicolumn{2}{c}{\textbf{Ours}} \\
\cmidrule(lr){4-6}
\cmidrule(lr){7-7}
\cmidrule(lr){8-8}
\cmidrule(lr){9-10}
& & & NR & NE & ND & BORF & GSR & RD-Sta & RD-Dyn \\
\midrule
\multirow{3}{*}{GN (0.25/0.02)}
& \(\rho=0.1\) 
& \(0.312(0.031)\) 
& \(0.284(0.034)\) 
& \(0.211(0.038)\) 
& \(0.330(0.036)\) 
& \(0.418(0.036)\) 
& \(0.267(0.056)\) 
& \(\underline{0.199(0.039)}\) 
& \(\mathbf{0.133(0.020)}\) \\

& \(\rho=0.2\) 
& \(0.382(0.031)\) 
& \(0.355(0.036)\) 
& \(0.292(0.046)\) 
& \(0.402(0.037)\) 
& \(0.492(0.036)\) 
& \(0.366(0.072)\) 
& \(\underline{0.285(0.049)}\) 
& \(\mathbf{0.170(0.030)}\) \\

& \(\rho=0.3\) 
& \(0.461(0.031)\) 
& \(0.440(0.045)\) 
& \(0.405(0.063)\) 
& \(0.480(0.043)\) 
& \(0.578(0.044)\) 
& \(0.453(0.072)\) 
& \(\underline{0.397(0.065)}\) 
& \(\mathbf{0.240(0.051)}\) \\
\midrule
\multirow{3}{*}{GN (0.25/0.05)}
& \(\rho=0.1\) 
& \(0.701(0.057)\) 
& \(0.671(0.059)\) 
& \(0.684(0.076)\) 
& \(0.723(0.060)\) 
& \(0.800(0.054)\) 
& \(0.561(0.069)\) 
& \(\underline{0.436(0.075)}\) 
& \(\mathbf{0.315(0.078)}\) \\

& \(\rho=0.2\) 
& \(0.792(0.061)\) 
& \(0.767(0.063)\) 
& \(0.812(0.081)\) 
& \(0.815(0.063)\) 
& \(0.890(0.061)\) 
& \(0.669(0.084)\) 
& \(\underline{0.568(0.085)}\) 
& \(\mathbf{0.455(0.087)}\) \\

& \(\rho=0.3\) 
& \(0.893(0.063)\) 
& \(0.874(0.067)\) 
& \(0.940(0.096)\) 
& \(0.914(0.065)\) 
& \(0.989(0.068)\) 
& \(0.805(0.103)\) 
& \(\underline{0.708(0.090)}\) 
& \(\mathbf{0.605(0.100)}\) \\
\midrule
\multirow{3}{*}{LFR}
& \(\mu=0.1\) 
& \(0.177(0.003)\) 
& \(0.098(0.006)\) 
& \(0.083(0.005)\) 
& \(0.233(0.005)\) 
& \(0.182(0.003)\) 
& \(0.135(0.002)\) 
& \(\underline{0.046(0.010)}\) 
& \(\mathbf{0.010(0.003)}\) \\

& \(\mu=0.2\) 
& \(0.412(0.004)\) 
& \(\underline{0.268(0.013)}\) 
& \(0.306(0.017)\) 
& \(0.553(0.012)\) 
& \(0.419(0.004)\) 
& \(0.304(0.003)\) 
& \(0.280(0.020)\) 
& \(\mathbf{0.017(0.005)}\) \\

& \(\mu=0.3\) 
& \(0.760(0.008)\) 
& \(\underline{0.583(0.020)}\) 
& \(0.721(0.023)\) 
& \(1.041(0.033)\) 
& \(0.770(0.008)\) 
& \(0.589(0.043)\) 
& \(0.708(0.025)\) 
& \(\mathbf{0.219(0.038)}\) \\
\bottomrule
\end{tabular}
}
\end{table*}

\subsection{Controlled Synthetic Denoising and Downstream Validation}

To complement the real-world evaluations, we conduct controlled experiments on synthetic graphs (GN \cite{girvan2002community} and LFR \cite{lancichinetti2008benchmark}) with known community structure. We evaluate explicit denoising quality using inverse signal-to-noise ratio (inv-SNR) \cite{yu2023network}, where lower values indicate better separation between intra- and inter-community edges. We further assess downstream GCN classification on the refined graphs.

Table~\ref{tab:synthetic_inv_snr} shows that both RD variants achieve strong explicit denoising performance. RD-Dyn obtains the best results across all settings, while RD-Sta is usually the second-best method. For example, on GN \((0.25/0.02)\) with \(\rho=0.2\), RD-Dyn reduces inv-SNR from \(0.382\) on the raw graph to \(0.170\), outperforming BORF (\(0.492\)) and GSR (\(0.366\)). On LFR with \(\mu=0.2\), RD-Dyn further reduces inv-SNR to \(0.017\), compared with \(0.419\) for BORF and \(0.304\) for GSR. These results indicate that integrating curvature modulation into a stable diffusion process is effective for explicit structure recovery.

Downstream GCN results, reported in Appendix~\uppercase\expandafter{\romannumeral 2}, further show that RD-refined graphs can benefit node classification in suitable regimes, although GSR achieves the strongest overall classification performance due to its learning-oriented design. For example, on GN \((0.25/0.02)\) with \(\rho=0.2\), RD-Dyn achieves an accuracy of \(0.763\), outperforming Raw (\(0.620\)) and GSR (\(0.757\)). On LFR with \(\mu=0.1\), RD-Sta obtains \(0.934\), higher than Raw (\(0.850\)) and GSR (\(0.896\)).

\section{Conclusion}

In this work, we revisit the conventional diffusion paradigm for network denoising from a geometric perspective and propose Ricci-Diffusion, a curvature-guided graph diffusion method inspired by Ricci flow. Experiments across real-world and controlled synthetic settings demonstrate that Ricci-Diffusion improves explicit denoising quality and often benefits downstream tasks. Beyond empirical gains, Ricci-Diffusion exhibits Ricci-flow-like behavior during denoising, characterized by curvature-guided edge-weight evolution and progressive curvature homogenization. These observations provide geometric insight into how curvature-aware diffusion yields more coherent and structurally consistent networks. Overall, the results suggest that incorporating curvature into graph diffusion is effective for structure-oriented network denoising.

\bibliographystyle{IEEEtran}
\bibliography{ref_papers}

\end{document}

\title{Network Denoising Revisited: A Ricci-Flow-Inspired Graph Diffusion Method \\ (Supplementary Material)}

\author{Anonymous Authors}

\maketitle

\begin{abstract}

This supplementary material provides the technical and experimental details that support the main paper. It complements the proposed Ricci-Diffusion framework from four aspects: theoretical justification, experimental reproducibility, computational efficiency, and robustness analysis. Section~\ref{Proofs} provides the proofs of two theoretical results: curvature separates graph structures that common similarity-driven diffusion kernels fail to distinguish, and curvature induces a first-order correction in one-step Ricci-Diffusion updates. Section~\ref{Protocols} describes the experimental protocols, covering biological network denoising, geometric validation, controlled synthetic structure recovery, and downstream GCN evaluation. Section~\ref{Computational} analyzes computational cost and scalability, focusing on curvature computation, runtime and memory usage. Section~\ref{Hyperparameter} reports hyperparameter sensitivity results, showing that Ricci-Diffusion is stable under reasonable parameter variations and that its improvement is mainly driven by curvature-aware diffusion.

\end{abstract}

\section{Proofs for Ricci-Diffusion}
\label{Proofs}

\subsection{Similarity-kernel Limitation and Curvature Separation}
The constructed family is unweighted, and ORC is evaluated under the unweighted shortest-path metric.

\begin{theorem}[Similarity-kernel limitation and curvature separation]
There exists a graph family \(\mathcal{M}=\{G_k\}_{k\ge 1}\) with the following property. For any \(G_k\in\mathcal{M}\), let \(A\) be its adjacency matrix, let \(D=\mathrm{diag}(A\mathbf 1)\), and set \(P=D^{-1}A\). Let \(D_P=\mathrm{diag}(P^\top\mathbf 1)\). Denote
\[
\mathcal{T}_{\rm sim}
=
\left\{
P,\;
D^{-1/2}AD^{-1/2},\;
PD_P^{-1}P^\top
\right\}.
\]
Then \( \exists u,a,v\in V(G_k)\) such that
\[
(u,a),(u,v)\in E(G_k),
\]
satisfying
\[
(T_{\rm sim})_{ua}=(T_{\rm sim})_{uv},
\quad
\forall T_{\rm sim}\in\mathcal{T}_{\rm sim},
\]
but
\[
\kappa_{ua}-\kappa_{uv}
=
\frac{2k}{k+3}
>0.
\]
\end{theorem}

\begin{proof}
For each \(k\ge 1\), we construct an unweighted graph
\(G_k=(V_k,E_k)\) as follows. Let
\[
V_k
=
\{u,a,v,c\}
\cup
\{x_i\}_{i=1}^{k}
\cup
\{y_i\}_{i=1}^{k+1}
\cup
\{z_i\}_{i=1}^{k+1}.
\]
The edge set \(E_k\) consists of
\[
(u,a),\ (u,v),\ (u,c),
\]
\[
(u,x_i),\qquad i=1,\ldots,k,
\]
\[
(a,c),\qquad (a,y_i),\qquad i=1,\ldots,k+1,
\]
\[
(v,c),\qquad (v,z_i),\qquad i=1,\ldots,k+1,
\]
and
\[
(x_i,y_i),\qquad i=1,\ldots,k.
\]
Let
\[
\mathcal{M}=\{G_k\}_{k\ge 1}.
\]
For every \(G_k\in\mathcal{M}\), the three nodes \(u,a,v\) satisfy
\[
(u,a),(u,v)\in E_k.
\]
Moreover,
\[
d_u=d_a=d_v=k+3.
\]

We first prove the indistinguishability under the three diffusion
kernels. Let \(A\) be the adjacency matrix of \(G_k\), let
\[
D=\operatorname{diag}(A\mathbf 1),
\qquad
P=D^{-1}A.
\]
For the random-walk kernel,
\[
T_{\rm rw}(u,a)
=
\frac{A_{ua}}{d_u}
=
\frac{1}{k+3}
=
\frac{A_{uv}}{d_u}
=
T_{\rm rw}(u,v).
\]
For the symmetric normalized kernel,
\[
T_{\rm sym}(u,a)
=
\frac{A_{ua}}{\sqrt{d_ud_a}}
=
\frac{1}{k+3},
\]
and
\[
T_{\rm sym}(u,v)
=
\frac{A_{uv}}{\sqrt{d_ud_v}}
=
\frac{1}{k+3}.
\]
Hence
\[
T_{\rm sym}(u,a)=T_{\rm sym}(u,v).
\]

It remains to verify the DSM kernel. By construction,
\[
N(u)\cap N(a)=\{c\},
\qquad
N(u)\cap N(v)=\{c\}.
\]
Thus
\[
T_{\rm dsm}(u,a)
=
\sum_{s\in V_k}
\frac{P_{us}P_{as}}{(P^\top\mathbf 1)_s}
=
\frac{P_{uc}P_{ac}}{(P^\top\mathbf 1)_c},
\]
and similarly
\[
T_{\rm dsm}(u,v)
=
\frac{P_{uc}P_{vc}}{(P^\top\mathbf 1)_c}.
\]
Since
\[
P_{uc}=P_{ac}=P_{vc}=\frac{1}{k+3},
\]
we obtain
\[
T_{\rm dsm}(u,a)=T_{\rm dsm}(u,v).
\]
Therefore,
\[
T_{\rm sim}(u,a)=T_{\rm sim}(u,v),
\qquad
\forall T_{\rm sim}\in\mathcal{T}_{\rm sim}.
\]

We now compute the Ollivier--Ricci curvature. Let \(\mu_p\) denote the
uniform one-hop probability measure at node \(p\), i.e.,
\[
\mu_p
=
\frac{1}{d_p}
\sum_{q\sim p}\delta_q.
\]
Since \(u\sim a\) and \(u\sim v\), under the unweighted shortest-path
metric,
\[
\kappa(u,a)=1-W_1(\mu_u,\mu_a),
\qquad
\kappa(u,v)=1-W_1(\mu_u,\mu_v).
\]
Set
\[
d=k+3.
\]

We first compute \(W_1(\mu_u,\mu_a)\). The neighborhoods are
\[
N(u)=\{a,v,c,x_1,\ldots,x_k\},
\]
and
\[
N(a)=\{u,c,y_1,\ldots,y_{k+1}\}.
\]
Consider the transport plan
\[
v\mapsto u,\qquad
c\mapsto c,\qquad
x_i\mapsto y_i\ (i=1,\ldots,k),
\qquad
a\mapsto y_{k+1}.
\]
The corresponding costs are
\[
1,\quad 0,\quad \underbrace{1,\ldots,1}_{k\ \text{times}},
\quad 1.
\]
Since each source has mass \(1/d\), this plan has total cost
\[
\frac{k+2}{d}.
\]
Hence
\[
W_1(\mu_u,\mu_a)\le \frac{k+2}{d}.
\]
Conversely, \(N(u)\cap N(a)=\{c\}\), so at most mass \(1/d\) can be
transported with zero cost. All remaining mass has transport cost at
least \(1\). Therefore,
\[
W_1(\mu_u,\mu_a)
\ge
1-\frac{1}{d}
=
\frac{k+2}{d}.
\]
Thus
\[
W_1(\mu_u,\mu_a)=\frac{k+2}{d},
\]
and consequently
\[
\kappa(u,a)
=
1-\frac{k+2}{d}
=
\frac{1}{d}
=
\frac{1}{k+3}.
\]

Next, we compute \(W_1(\mu_u,\mu_v)\). The neighborhoods are
\[
N(u)=\{a,v,c,x_1,\ldots,x_k\},
\]
and
\[
N(v)=\{u,c,z_1,\ldots,z_{k+1}\}.
\]
Consider the transport plan
\[
c\mapsto c,\qquad
a\mapsto u,\qquad
v\mapsto z_{k+1},
\qquad
x_i\mapsto z_i\ (i=1,\ldots,k).
\]
The corresponding costs are
\[
0,\quad 1,\quad 1,\quad
\underbrace{3,\ldots,3}_{k\ \text{times}}.
\]
Therefore,
\[
W_1(\mu_u,\mu_v)
\le
\frac{3k+2}{d}.
\]

We prove the reverse inequality. Let
\[
Z=\{z_1,\ldots,z_{k+1}\}.
\]
For an arbitrary coupling \(\pi\) from \(\mu_u\) to \(\mu_v\), define
\[
m_v=\sum_{z\in Z}\pi(v,z),
\qquad
m_c=\sum_{z\in Z}\pi(c,z).
\]
Since each source has mass \(1/d\),
\[
0\le m_v\le \frac{1}{d},
\qquad
0\le m_c\le \frac{1}{d}.
\]
The total target mass on \(Z\) is \((k+1)/d\). Moreover, the distances
from the source nodes in \(N(u)\) to \(Z\) satisfy
\[
d(v,Z)=1,\qquad d(c,Z)=2,\qquad
d(s,Z)\ge 3
\]
for all
\[
s\in \{a,x_1,\ldots,x_k\}.
\]
Hence the cost of transporting mass to \(Z\) is at least
\[
m_v+2m_c
+
3\left(
\frac{k+1}{d}-m_v-m_c
\right)
=
\frac{3(k+1)}{d}
-
2m_v
-
m_c.
\]
The target node \(u\) receives mass \(1/d\), and every source node in
\(N(u)\) has distance at least \(1\) to \(u\). Thus transporting mass to
\(u\) costs at least \(1/d\). In addition, if mass \(m_c\) is sent from
\(c\) to \(Z\), then at least mass \(m_c\) must be transported to the
target node \(c\) from sources other than \(c\), which costs at least
\(m_c\). Therefore, the cost of transporting mass to \(\{u,c\}\) is at
least
\[
\frac{1}{d}+m_c.
\]
Combining the two lower bounds gives
\[
W_1(\mu_u,\mu_v)
\ge
\frac{3(k+1)}{d}
-
2m_v
-
m_c
+
\frac{1}{d}
+
m_c
=
\frac{3k+4}{d}
-
2m_v.
\]
Since \(m_v\le 1/d\), we obtain
\[
W_1(\mu_u,\mu_v)
\ge
\frac{3k+2}{d}.
\]
Together with the feasible transport plan above,
\[
W_1(\mu_u,\mu_v)=\frac{3k+2}{d}.
\]
Thus
\[
\kappa(u,v)
=
1-\frac{3k+2}{d}
=
-\frac{2k-1}{d}
=
-\frac{2k-1}{k+3}.
\]
Therefore,
\[
\kappa(u,a)-\kappa(u,v)
=
\frac{1}{k+3}
+
\frac{2k-1}{k+3}
=
\frac{2k}{k+3}
>0.
\]
This completes the proof.
\end{proof}

\subsection{Curvature-Induced First-Order Correction}

\begin{proposition}[Curvature-induced first-order correction]
Let \(T_s\in\mathbb{R}_{\ge 0}^{n\times n}\).
Let \(\mathcal{N}_i=\{j:T_s(i,j)>0\}\), and define the parameterized curvature-aware kernel by
\[
T_\eta=T_\kappa(\eta)=\Pi_{\mathrm{DSM}}(A_\eta),
\]
where
\[
A_\eta(i,j)
=
\frac{e^{\eta\kappa_{ij}}T_s(i,j)}
{\sum_{\ell\in\mathcal{N}_i}e^{\eta\kappa_{i\ell}}T_s(i,\ell)}
\]
for \(j\in\mathcal{N}_i\), and \(A_\eta(i,j)=0\) otherwise.
Let \(T_0=T_\kappa(0)\). Define
\[
\bar{\kappa}_i
=
\sum_{\ell\in\mathcal{N}_i}T_s(i,\ell)\kappa_{i\ell},~
(B_\kappa)_{ij}
=
\begin{cases}
T_s(i,j)(\kappa_{ij}-\bar{\kappa}_i), & j\in\mathcal{N}_i,\\
0, & j\notin\mathcal{N}_i.
\end{cases}
\]
Let
\[
\dot T_0=\mathrm{D}\Pi_{\mathrm{DSM}}[T_s](B_\kappa),
\]
where \(\mathrm{D}\Pi_{\mathrm{DSM}}[T_s]\) denotes the Fréchet derivative of \(\Pi_{\mathrm{DSM}}\) at \(T_s\).
For \(W\in\mathbb{R}_{\ge 0}^{n\times n}\) and \(\tau\in(0,1)\), define
\[
W_\eta^+
=
\tau T_\eta W T_\eta+(1-\tau)T_\eta,
\quad
W_{\eta=0}^+
=
\tau T_0WT_0+(1-\tau)T_0.
\]
Then, as \(\eta\to 0\),
\[
W_\eta^+
=
W_{\eta=0}^+
+
\eta\mathcal{G}_\kappa(W,T_0)
+
O(\eta^2),
\]
where
\[
\mathcal{G}_\kappa(W,T_0)
=
(1-\tau)\dot T_0
+
\tau
\left(
\dot T_0WT_0+T_0W\dot T_0
\right).
\]
\end{proposition}

\begin{proof}
For each \(i\), define
\[
Z_i(\eta)
=
\sum_{\ell\in\mathcal{N}_i}
e^{\eta\kappa_{i\ell}}T_s(i,\ell).
\]
Then, for \(j\in\mathcal{N}_i\),
\[
A_\eta(i,j)
=
\frac{e^{\eta\kappa_{ij}}T_s(i,j)}{Z_i(\eta)}.
\]
Since \(T_s\) is row-stochastic, we have
\[
Z_i(0)
=
\sum_{\ell\in\mathcal{N}_i}T_s(i,\ell)
=
1,
\]
and hence
\[
A_0(i,j)=T_s(i,j).
\]

We first compute the first-order variation of \(A_\eta\) at \(\eta=0\).
For \(j\in\mathcal{N}_i\),
\[
\log A_\eta(i,j)
=
\eta\kappa_{ij}
+
\log T_s(i,j)
-
\log Z_i(\eta).
\]
Differentiating with respect to \(\eta\), we obtain
\[
\frac{\partial}{\partial\eta}\log A_\eta(i,j)
=
\kappa_{ij}
-
\frac{Z_i'(\eta)}{Z_i(\eta)}.
\]
Moreover,
\[
Z_i'(\eta)
=
\sum_{\ell\in\mathcal{N}_i}
\kappa_{i\ell}e^{\eta\kappa_{i\ell}}T_s(i,\ell).
\]
Evaluating at \(\eta=0\) gives
\[
\frac{Z_i'(0)}{Z_i(0)}
=
\sum_{\ell\in\mathcal{N}_i}T_s(i,\ell)\kappa_{i\ell}
=
\bar{\kappa}_i.
\]
Therefore,
\[
\left.
\frac{\partial}{\partial\eta}\log A_\eta(i,j)
\right|_{\eta=0}
=
\kappa_{ij}-\bar{\kappa}_i.
\]
Since
\[
\frac{\partial A_\eta(i,j)}{\partial\eta}
=
A_\eta(i,j)
\frac{\partial}{\partial\eta}\log A_\eta(i,j),
\]
we have
\[
\left.
\frac{\partial A_\eta(i,j)}{\partial\eta}
\right|_{\eta=0}
=
T_s(i,j)(\kappa_{ij}-\bar{\kappa}_i)
=
(B_\kappa)_{ij}.
\]
For \(j\notin\mathcal{N}_i\), both \(A_\eta(i,j)\) and
\((B_\kappa)_{ij}\) are zero. Hence, in matrix form,
\[
A_\eta
=
T_s+\eta B_\kappa+O(\eta^2).
\]

By the Fréchet differentiability of \(\Pi_{\mathrm{DSM}}\) at \(T_s\),
we obtain
\[
T_\eta
=
\Pi_{\mathrm{DSM}}(A_\eta)
=
\Pi_{\mathrm{DSM}}(T_s)
+
\eta
\mathrm{D}\Pi_{\mathrm{DSM}}[T_s](B_\kappa)
+
O(\eta^2).
\]
Since
\[
T_0=\Pi_{\mathrm{DSM}}(T_s),
\qquad
\dot T_0
=
\mathrm{D}\Pi_{\mathrm{DSM}}[T_s](B_\kappa),
\]
it follows that
\[
T_\eta
=
T_0+\eta\dot T_0+O(\eta^2).
\]
Let
\[
R_\eta=T_\eta-T_0.
\]
Then
\[
R_\eta=\eta\dot T_0+O(\eta^2).
\]

Using \(T_\eta=T_0+R_\eta\), we expand the one-step update:
\begin{align}
W_\eta^+
&=
\tau (T_0+R_\eta)W(T_0+R_\eta)
+
(1-\tau)(T_0+R_\eta) \notag\\
&=
W_{\eta=0}^+
+
(1-\tau)R_\eta \notag\\
&\quad
+
\tau R_\eta WT_0
+
\tau T_0WR_\eta
+
\tau R_\eta WR_\eta .
\end{align}
Substituting
\[
R_\eta=\eta\dot T_0+O(\eta^2)
\]
into the above identity gives
\[
(1-\tau)R_\eta
=
\eta(1-\tau)\dot T_0+O(\eta^2),
\]
\[
\tau R_\eta WT_0
=
\eta\tau\dot T_0WT_0+O(\eta^2),
\]
\[
\tau T_0WR_\eta
=
\eta\tau T_0W\dot T_0+O(\eta^2),
\]
and
\[
\tau R_\eta WR_\eta
=
O(\eta^2).
\]
Therefore,
\[
W_\eta^+
=
W_{\eta=0}^+
+
\eta
\left[
(1-\tau)\dot T_0
+
\tau
\left(
\dot T_0WT_0+T_0W\dot T_0
\right)
\right]
+
O(\eta^2).
\]
By the definition of \(\mathcal{G}_\kappa(W,T_0)\), this is exactly
\[
W_\eta^+
=
W_{\eta=0}^+
+
\eta\mathcal{G}_\kappa(W,T_0)
+
O(\eta^2).
\]
The proof is complete.
\end{proof}

\section{Experimental Protocols}
\label{Protocols}

This section provides additional implementation details for the experiments in the main paper. Unless otherwise specified, all methods are evaluated under the same input graphs and downstream evaluation protocols within each task. The best result is highlighted in bold in the main paper, and the second-best result is underlined.

\subsection{Common Setup}

For real-world denoising tasks, Ricci-Diffusion is compared with representative diffusion-based network denoising methods, including Network Deconvolution (ND), Network Enhancement (NE), and Network Refinement (NR). For controlled synthetic experiments, we additionally include BORF and GSR. BORF is used as a curvature-based graph rewiring baseline, while GSR is included as a learning-oriented graph structure refinement baseline for downstream GCN evaluation.

Ricci-Diffusion is evaluated in two variants. RD-Sta computes the curvature-aware diffusion kernel once and keeps it fixed during the diffusion process. RD-Dyn adopts the dynamic-to-static strategy, where the curvature-aware kernel is recomputed during a short dynamic stage and then fixed in the subsequent static diffusion stage. Unless otherwise specified, we use \(\eta=1\), \(\tau=0.9\), \(k=20\), \(q=1\), and \(T_{\mathrm{dyn}}=2\).

\subsection{Tissue Gene-Function Prediction}

We evaluate network denoising on tissue-specific gene interaction networks \cite{greene2015understanding} through gene function prediction. The tissue networks are constructed following the protocol of NE \cite{wang2018network}, yielding 16 tissues with valid evaluation functions. After applying each denoising method to the tissue-specific network, gene-function prediction is performed using weighted random walks with restart \cite{kohler2008walking} on the denoised graph.

Performance is evaluated by leave-one-out cross-validation against experimentally validated gene-function associations. The evaluation metric is the area under the receiver operating characteristic curve (AUROC). Higher AUROC indicates that the denoised network better supports the recovery of biologically meaningful gene-function relationships.

\subsection{Hi-C TAD Detection}

We evaluate whether network denoising improves topologically associating domain (TAD) detection from Hi-C contact networks \cite{rao20143d, schmitt2016genome}. The experiments use Hi-C data from the GM12878 cell line at 1\,k and 5\,k resolutions, considering all autosomes. For each chromosome-level contact network, a denoising method is first applied to the raw contact matrix.

After denoising, Louvain community detection \cite{blondel2008fast} is applied to the resulting network. The detected communities are compared with the reference domain structure using normalized mutual information (NMI). Higher NMI indicates better agreement between the detected communities and the reference TAD structure.

\subsection{Butterfly Retrieval}

We evaluate fine-grained species identification on the Leeds Butterfly dataset \cite{wang2009learning}, which contains 832 images from 10 species. Each image is represented using Fisher Vector \cite{sanchez2011fisher} and VLAD \cite{herve2010aggregating} descriptors extracted from dense SIFT features \cite{bosch2007image}. The corresponding image similarity networks are constructed from inner-product similarities and used as the input graphs for denoising.

The denoised network is evaluated through an image retrieval task. For each query image, all other images are ranked according to their similarities in the denoised network. Retrieval accuracy is computed as the proportion of same-species images among the top-ranked neighbors and then averaged over all query images.

\subsection{Geometric Validation}

To examine Ricci-flow-like behavior, we analyze the coupling between edge-weight evolution and Ollivier--Ricci curvature along the denoising trajectory. This experiment is conducted on the Leeds Butterfly dataset using a single run with \(T=40\) iterations. At each iteration \(t\), we record the edge-weight matrix \(W^{(t)}\) and recompute the corresponding curvature matrix \(\kappa^{(t)}\).

For each edge, we measure the relative edge-weight change
\[
r^{(t)}_{ij}
=
\log w^{(t)}_{ij}
-
\log w^{(t+1)}_{ij}.
\]
We then compute Pearson and Spearman correlations between \(r^{(t)}_{ij}\) and \(\kappa^{(t)}_{ij}\) across iterations. The evolution of curvature distributions is also recorded to assess whether the denoising trajectory induces curvature concentration.

\subsection{Ablation Protocol}

The ablation study is conducted on the Leeds Butterfly dataset to isolate the effect of curvature modulation from the diffusion backbone. We vary the curvature signal used in the diffusion kernel, including Ollivier--Ricci curvature (ORC), Forman--Ricci curvature (FRC), random signals, and degree-based structural proxies. We also vary the curvature strength parameter \(\eta\).

The no-curvature setting is obtained by setting \(\eta=0\), which removes curvature modulation while keeping the same diffusion backbone. For RD-Dyn, we use \(q=1\) and \(T_{\mathrm{dyn}}=2\) unless otherwise specified. For RD-Sta, the curvature-aware kernel is computed once and then fixed throughout the diffusion process. The default diffusion and sparsification parameters are \(\tau=0.9\) and \(k=20\).

\begin{table*}[t]
\centering
\caption{Downstream GCN classification on synthetic graphs. The metric is accuracy. Higher is better.}
\label{tab:synthetic_gcn_acc}
\scriptsize
\setlength{\tabcolsep}{4.0pt}
\resizebox{\textwidth}{!}{
\begin{tabular}{cccccccccc}
\toprule
\multirow{2}{*}{\textbf{Dataset}}
& \multirow{2}{*}{\textbf{Noise}}
& \multirow{2}{*}{\textbf{Raw}}
& \multicolumn{3}{c}{\textbf{Diffusion baselines}} 
& \textbf{Curv.} 
& \textbf{Learn.} 
& \multicolumn{2}{c}{\textbf{Ours}} \\
\cmidrule(lr){4-6}
\cmidrule(lr){7-7}
\cmidrule(lr){8-8}
\cmidrule(lr){9-10}
& & & NR & NE & ND & BORF & GSR & RD-Sta & RD-Dyn \\
\midrule
\multirow{3}{*}{GN (0.25/0.02)}
& \(\rho=0.1\) 
& \(0.668(0.132)\) 
& \(0.723(0.122)\) 
& \(0.745(0.103)\) 
& \(0.650(0.153)\) 
& \(0.584(0.110)\) 
& \(\mathbf{0.789(0.133)}\) 
& \(\underline{0.752(0.137)}\) 
& \(0.739(0.167)\) \\

& \(\rho=0.2\) 
& \(0.620(0.118)\) 
& \(0.691(0.122)\) 
& \(0.739(0.089)\) 
& \(0.632(0.109)\) 
& \(0.538(0.110)\) 
& \(\underline{0.757(0.112)}\) 
& \(0.711(0.112)\) 
& \(\mathbf{0.763(0.150)}\) \\

& \(\rho=0.3\) 
& \(0.546(0.109)\) 
& \(0.611(0.106)\) 
& \(0.643(0.140)\) 
& \(0.538(0.122)\) 
& \(0.509(0.097)\) 
& \(\mathbf{0.696(0.136)}\) 
& \(0.652(0.111)\) 
& \(\underline{0.673(0.133)}\) \\
\midrule
\multirow{3}{*}{GN (0.25/0.05)}
& \(\rho=0.1\) 
& \(0.500(0.090)\) 
& \(0.554(0.117)\) 
& \(0.550(0.133)\) 
& \(0.484(0.113)\) 
& \(0.430(0.086)\) 
& \(\mathbf{0.655(0.121)}\) 
& \(\underline{0.588(0.117)}\) 
& \(0.486(0.166)\) \\

& \(\rho=0.2\) 
& \(0.489(0.098)\) 
& \(0.521(0.116)\) 
& \(\underline{0.548(0.100)}\) 
& \(0.445(0.117)\) 
& \(0.427(0.098)\) 
& \(\mathbf{0.609(0.092)}\) 
& \(0.527(0.133)\) 
& \(0.452(0.139)\) \\

& \(\rho=0.3\) 
& \(0.471(0.100)\) 
& \(0.493(0.093)\) 
& \(0.446(0.110)\) 
& \(0.409(0.127)\) 
& \(0.395(0.093)\) 
& \(\mathbf{0.496(0.097)}\) 
& \(\underline{0.494(0.138)}\) 
& \(0.382(0.106)\) \\
\midrule
\multirow{3}{*}{LFR}
& \(\mu=0.1\) 
& \(0.850(0.046)\) 
& \(0.901(0.051)\) 
& \(\underline{0.910(0.053)}\) 
& \(0.770(0.033)\) 
& \(0.833(0.055)\) 
& \(0.896(0.041)\) 
& \(\mathbf{0.934(0.046)}\) 
& \(0.707(0.083)\) \\

& \(\mu=0.2\) 
& \(0.668(0.048)\) 
& \(\underline{0.746(0.046)}\) 
& \(0.605(0.043)\) 
& \(0.441(0.044)\) 
& \(0.641(0.045)\) 
& \(\mathbf{0.811(0.047)}\) 
& \(0.615(0.080)\) 
& \(0.604(0.076)\) \\

& \(\mu=0.3\) 
& \(0.478(0.029)\) 
& \(\underline{0.551(0.050)}\) 
& \(0.334(0.045)\) 
& \(0.260(0.054)\) 
& \(0.467(0.041)\) 
& \(\mathbf{0.626(0.051)}\) 
& \(0.338(0.059)\) 
& \(0.358(0.060)\) \\
\bottomrule
\end{tabular}
}
\end{table*}

\subsection{Synthetic Graphs and GCN Evaluation}

We use controlled synthetic graphs with known community structure to evaluate explicit structure recovery. For GN graphs \cite{girvan2002community}, we generate graphs with 128 nodes and four equal-sized communities. We consider two community-separation regimes, \(p_{\mathrm{in}}/p_{\mathrm{out}}=0.25/0.02\) and \(0.25/0.05\). Structural noise is injected by removing intra-community edges with probability \(\rho/2\) and adding approximately the same number of random inter-community edges, with \(\rho\in\{0.1,0.2,0.3\}\).

For LFR graphs \cite{lancichinetti2008benchmark}, we generate graphs with 1000 nodes and vary the mixing parameter \(\mu\in\{0.1,0.2,0.3\}\), where larger \(\mu\) indicates weaker community separability. Explicit denoising quality is measured by inverse signal-to-noise ratio (inv-SNR), defined as the ratio between inter-community and intra-community edge weights. Lower inv-SNR indicates better denoising. 

For downstream validation, we train the same two-layer GCN on each refined graph. Ground-truth communities are used as node labels, and all methods share the same train/validation/test split. To keep the comparison consistent with the structure-only setting, we use only intrinsic structural node features, including degree, PageRank, and local clustering coefficient, without introducing external information. The results are presented in Table~\ref{tab:synthetic_gcn_acc}.

\section{Computational Cost and Scalability}
\label{Computational}

We analyze the computational cost of Ricci-Diffusion (RD) from both theoretical and empirical perspectives. The algorithm consists of four main components: graph sparsification, construction of the local transition field, curvature computation, and iterative diffusion. Among them, the dominant computational bottleneck is the computation of Ollivier--Ricci curvature (ORC), which requires solving local optimal transport problems over neighborhood supports.

Let \(n=|V|\), \(m=|E|\), and let \(D\) denote the maximum neighborhood size after sparsification. Since ORC is computed edge-wise, the total cost depends on solving a local OT problem for each retained edge. Exact solvers such as network simplex may incur \(O(D^3)\) cost on each local support. In our implementation, we instead use Sinkhorn iterations, whose per-iteration cost is \(O(D^2)\). Thus the curvature computation cost is
\[
O\bigl(m\,S\,D^2\bigr),
\]
where \(S\) is the number of Sinkhorn iterations; with fixed \(S\), this becomes \(O(mD^2)\). Since RD applies \(k\)NN sparsification, \(D\) is controlled by \(k\), making the curvature computation local rather than fully dense.

The remaining diffusion update
\[
W^{(t+1)}=\tau T_{\kappa}W^{(t)}T_{\kappa}+(1-\tau)T_{\kappa}
\]
mainly involves sparse matrix multiplications when \(T_{\kappa}\) is sparsified. Therefore, compared with standard diffusion-based methods, the additional overhead of RD primarily comes from curvature estimation rather than from the diffusion iteration itself.

RD provides two variants with different cost--performance trade-offs. RD-Dyn periodically recomputes the curvature and the associated diffusion kernel during the dynamic stage, allowing the transport geometry to adapt to the evolving graph. This usually improves denoising quality but increases runtime. RD-Sta computes the curvature-aware kernel once and then reuses it throughout the static diffusion stage, thereby avoiding repeated curvature computation and significantly reducing the overall cost. This design gives users a flexible choice between stronger performance and lower computational overhead.

We report runtime and memory usage for all methods under the same protocol. Runtime is averaged over three runs excluding data loading time, and memory usage is measured as the additional peak memory consumed by the algorithm process. All experiments are conducted on a machine with an Intel i7-12700K CPU and 32GB RAM.

\begin{table}[h]
\centering
\caption{Runtime on the Butterfly dataset.}
\label{tab:butterfly_runtime}
\begin{tabular}{lccccc}
\toprule
Method & RD-Dyn & RD-Sta & NR & NE & ND \\
\midrule
Time (s) & 4.6778 & 1.2783 & 1.6397 & 0.1182 & 0.1479 \\
\bottomrule
\end{tabular}
\end{table}

Table~\ref{tab:butterfly_runtime} reports the runtime on the Butterfly dataset. RD-Sta reduces the cost of RD-Dyn while retaining the curvature-aware diffusion mechanism.

\begin{table}[h]
\centering
\caption{Runtime scaling on GN synthetic graphs.}
\label{tab:runtime_scaling}
\begin{tabular}{lccccc}
\toprule
Size & RD-Dyn & RD-Sta & NR & NE & ND \\
\midrule
100  & 0.8862   & 0.7607   & 0.1767  & 0.0063  & 0.0432 \\
500  & 3.5572   & 1.2579   & 0.6649  & 0.0874  & 0.2483 \\
1000 & 10.6259  & 4.4282   & 1.6580  & 0.3593  & 0.2563 \\
2000 & 37.6926  & 15.7429  & 6.2109  & 1.4949  & 1.1519 \\
5000 & 272.8133 & 108.5649 & 41.9042 & 16.7649 & 14.3154 \\
\bottomrule
\end{tabular}
\end{table}

Table~\ref{tab:runtime_scaling} shows the runtime scaling on GN synthetic graphs. Across graph sizes, RD-Sta consistently reduces runtime compared with RD-Dyn, demonstrating the practical effect of the dynamic-to-static strategy.

\begin{table}[h]
\centering
\caption{Memory scaling on GN synthetic graphs.}
\label{tab:memory_scaling}
\begin{tabular}{lccccc}
\toprule
Size & RD-Dyn & RD-Sta & NR & NE & ND \\
\midrule
100  & 523.715  & 523.723  & 265.348  & 325.922  & 1226.453 \\
500  & 607.402  & 580.340  & 303.375  & 349.105  & 1275.289 \\
1000 & 823.984  & 755.688  & 353.312  & 417.953  & 1510.695 \\
2000 & 1496.488 & 1392.900 & 573.600  & 643.300  & 1918.100 \\
5000 & 4483.200 & 4489.400 & 2260.300 & 2358.600 & 4076.600 \\
\bottomrule
\end{tabular}
\end{table}

Table~\ref{tab:memory_scaling} reports the memory usage on GN synthetic graphs. The memory growth is stable with graph size and remains comparable to other matrix-based denoising methods at larger scales.

Overall, these results show that RD provides a controllable cost--performance trade-off through its dynamic and static variants. Since curvature computation is a modular component, the implementation can be further accelerated using approximate optimal transport, batched edge-wise computation, parallelization, or lower-cost curvature surrogates such as Forman--Ricci curvature.

\section{Hyperparameter Sensitivity Analysis}
\label{Hyperparameter}

We evaluate the sensitivity of Ricci-Diffusion on the Butterfly dataset using mean retrieval accuracy, where higher values indicate better denoising quality. Unless otherwise specified, the default setting is \(\eta=1\), \(\tau=0.9\), \(k=20\), and \(q=1,T_{\mathrm{dyn}}=2\) for RD-Dyn.

\begin{table}[h]
\centering
\caption{Ablation and sensitivity analysis on the Butterfly dataset. Higher is better.}
\label{tab:ablation_sensitivity}
\scriptsize
\setlength{\tabcolsep}{4pt}
\begin{tabular}{llcc}
\toprule
Group & Setting & RD-Dyn & RD-Sta \\
\midrule
\multicolumn{4}{l}{\textit{Curvature strength and type}} \\
No curvature & \(\eta=0\)       & 0.789 & 0.759 \\
ORC          & \(\eta=1\)       & 0.817 & 0.783 \\
ORC          & \(\eta=2\)       & 0.824 & 0.789 \\
ORC          & \(\eta=3\)       & 0.825 & 0.794 \\
FRC          & \(\eta=2\)       & 0.806 & 0.775 \\
Random       & --               & 0.573 & 0.556 \\
Degree       & --               & 0.709 & 0.691 \\
\midrule
\multicolumn{4}{l}{\textit{Sparsification and diffusion strength}} \\
\(k\)        & 5                & 0.733 & 0.758 \\
\(k\)        & 10               & 0.789 & 0.781 \\
\(k\)        & 15               & 0.810 & 0.784 \\
\(k\)        & 20               & 0.817 & 0.783 \\
\(k\)        & 25               & 0.810 & 0.778 \\
\(\tau\)    & 0.1              & 0.703 & 0.679 \\
\(\tau\)    & 0.3              & 0.735 & 0.707 \\
\(\tau\)    & 0.5              & 0.759 & 0.731 \\
\(\tau\)    & 0.7              & 0.783 & 0.755 \\
\(\tau\)    & 0.9              & 0.817 & 0.783 \\
\bottomrule
\end{tabular}
\end{table}

Table~\ref{tab:ablation_sensitivity} summarizes the main ablation and sensitivity results. Removing curvature leads to lower performance, while ORC consistently improves both RD-Dyn and RD-Sta. FRC remains competitive, whereas random and degree-based replacements are clearly weaker, indicating that the gain comes from structured edge-level geometric information. The results are also stable across reasonable choices of \(k\) and \(\tau\), with stronger diffusion and moderately large neighborhoods giving better performance.

For the dynamic schedule, we vary \(q\in\{1,5,10,20\}\) and \(T_{\mathrm{dyn}}\in\{1,2,3,4\}\). RD-Dyn remains generally stable across these settings, with ORC-based accuracies ranging from 0.755 to 0.835; the default short schedule \(q=1,T_{\mathrm{dyn}}=2\) already achieves 0.817.

Overall, these results show that RD does not rely on fragile hyperparameter tuning. Its improvement is mainly driven by curvature-aware diffusion, and the default configuration provides a stable balance between performance and simplicity.

\paragraph{Summary.}
Overall, RD is robust to reasonable hyperparameter variations. The most important factor is whether curvature modulation is used: removing curvature degrades performance, while replacing ORC with random or degree-based signals fails to recover the gain. The remaining parameters mainly control the strength and locality of the diffusion process. Performance is stable for moderate-to-large \(k\), improves with stronger diffusion \(\tau\), and remains robust under a range of dynamic schedules. These results indicate that the improvement of RD is not due to fragile hyperparameter tuning, but to the coupling between curvature-aware modulation and diffusion.

\bibliographystyle{IEEEtran}
\bibliography{ref_papers}